\documentclass[12pt]{article}

\pdfoutput=1
\usepackage{color}
\usepackage{epsfig, palatino}
\usepackage{pstricks,pst-node,pst-tree}
\usepackage{epic}
\usepackage{mathrsfs}
\usepackage{ae} 
\usepackage[T1]{fontenc}
\usepackage[ansinew]{inputenc}
\usepackage{amsmath}
\usepackage{amssymb}
\usepackage{graphicx}
\usepackage{ulem}
\usepackage{color}
\definecolor{darkblue}{cmyk}{0.9,0.9,0,0}
\usepackage[colorlinks=true,linkcolor=darkblue,citecolor=darkblue,urlcolor=darkblue]{hyperref}
\usepackage{cite}
\usepackage{hyperref}
\usepackage{wasysym}
\usepackage{varioref}
\usepackage{makeidx}
\usepackage[english]{babel}
\usepackage{simplewick}
\usepackage{array}
\usepackage{multirow}

\usepackage[font={small}]{caption}

\newcommand{\comment}[1]{}

\newcommand{\beq}{\begin{equation}}
\newcommand{\eeq}{\end{equation}}
\newcommand{\beqq}{\begin{equation*}}
\newcommand{\eeqq}{\end{equation*}}
\newcommand\beqa{\begin{eqnarray}}
\newcommand\eeqa{\end{eqnarray}}
\newcommand\beqaa{\begin{eqnarray*}}
\newcommand\eeqaa{\end{eqnarray*}}
\newcommand\bea{\begin{array}}
\newcommand\eea{\end{array}}

\newcommand{\nn}{\nonumber}

\newcommand{\neqa}{\nonumber\end{eqnarray}} 
\newcommand{\la}[1]{\label{#1}}

\renewcommand{\d}{\partial}

\newcommand{\<}{{\langle}}
\renewcommand{\>}{{\rangle}}

\newcommand{\re}{\relax{\rm I\kern-.18em R}}

\renewcommand{\sp}{p\hspace{-.40em}/}

\definecolor{darkgreen}{rgb}{0.0, 0.45, 0.0}

\def\XXint#1#2#3{{\setbox0=\hbox{$#1{#2#3}{\int}$}
\vcenter{\hbox{$#2#3$}}\kern-.5\wd0}}

\def\su2{{SU(2)}}

\def\[{\left[}
\def\]{\right]}

\def\({\left(}
\def\){\right)}
\def\[{\left[}
\def\]{\right]}

\def\<{\langle}
\def\>{\rangle}

\def\i2{\frac{i}{2}}

\def\spi{\relax{\rm \pi\kern-0.5em /}}
\def\sA{\relax{\rm A\kern-0.5em /}}
\def\sp{\relax{\rm p\kern-0.5em /}}
\def\sd{\relax{\rm \d\kern-0.5em /}}
\def\sk{\relax{\rm k\kern-0.5em /}}
\def\sn{\relax{\rm n\kern-0.5em /}}
\def\sl{\relax{\rm l\kern-0.5em /}}
\def\sP{\relax{\rm P\kern-0.7em /}}
\def\sBethe{\relax{\rm \Bethe\kern-0.5em /}}

\def\2F1{\,_2{\rm F}_1}

\def\wB{{\bf w}_{\bf B}}
\def\wL{{\bf w}_{\bf L}}
\def\wR{{\bf w}_{\bf R}}
\def\bottom#1{\textcolor[rgb]{1,0,0}{#1}}
\def\lefta#1{\textcolor[rgb]{0,0,1}{#1}}
\def\righta#1{\textcolor[rgb]{0,0.6,0}{#1}}

\makeindex

\begin{document}

\thispagestyle{empty}

\renewcommand{\thefootnote}{\fnsymbol{footnote}}
\setcounter{page}{1}
\setcounter{footnote}{0}
\setcounter{figure}{0}

\begin{center}
$$$$
{\Large\textbf{\mathversion{bold}
Gluing Hexagons at Three Loops
}\par}

\vspace{1.0cm}

\textrm{Benjamin Basso$^\text{\tiny 1}$, Vasco Goncalves$^\text{\tiny 2}$, Shota Komatsu$^\text{\tiny 3}$, Pedro Vieira$^\text{\tiny 3}$}
\\ \vspace{1.2cm}
\footnotesize{\textit{
$^\text{\tiny 1}$Laboratoire de Physique Th\'eorique, \'Ecole Normale Sup\'erieure, Paris 75005, France\\
$^\text{\tiny 2}$Centro de F$\acute{\imath}$sica do Porto, Departamento de F$\acute{\imath}$sica e Astronomia,
Faculdade de Ci$\hat{e}$ncias da Universidade do Porto, Rua do Campo Alegre 687, 4169-007 Porto, Portugal\\
$^\text{\tiny 3}$Perimeter Institute for Theoretical Physics,
Waterloo, Ontario N2L 2Y5, Canada
}  
\vspace{4mm}
}

\par\vspace{1.5cm}

\textbf{Abstract}\vspace{2mm}
\end{center}
We perform extensive three-loop tests of the hexagon bootstrap approach for structure constants in planar $\mathcal{N}=4$ SYM theory. We focus on correlators involving two BPS operators and one non-BPS operator in the so-called $SL(2)$ sector. At three loops, such correlators receive wrapping corrections from mirror excitations flowing in either the adjacent or the opposing channel. 
Amusingly, we find that the first type of correction coincides exactly with the leading wrapping correction for the spectrum (divided by the one-loop anomalous dimension). We develop an efficient method for computing the second type of correction for operators with any spin. The results are in perfect agreement with the recently obtained three-loop perturbative data by Chicherin, Drummond, Heslop, Sokatchev \cite{Emery} and by Eden \cite{Eden}. We also derive the integrand for general multi-particle wrapping corrections, which turns out to take a remarkably simple form. As an application we estimate the loop order at which various new physical effects are expected to kick-in.

\noindent

\setcounter{page}{1}
\renewcommand{\thefootnote}{\arabic{footnote}}
\setcounter{footnote}{0}

\setcounter{tocdepth}{2}

 \def\nref#1{{(\ref{#1})}}

\newpage

\tableofcontents

\parskip 5pt plus 1pt   \jot = 1.5ex

\section{Introduction} 
In \cite{C123Paper} a simple proposal for studying 3-point correlation functions in planar $\mathcal{N}=4$ SYM was put forward. It is a sort of \textit{divide and conquer} strategy where the 3-point correlator -- represented as the usual string pair of paints -- is cut  {into} two simpler hexagonal building blocks which are bootstrapped using integrability and then stitched back together. The  {cutting procedure} involves summing over partitions of the rapidities of the physical particles while the stitching back  {together requires} integrating over the rapidities of the mirror particles, see figure \ref{hexagon}. The leading process with  {no mirror particle exchanged} is called the \textit{asymptotic} result while processes with mirror excitations  travelling  {around}  are referred to as \textit{wrapping effects}.

In this paper we present a series of tests for the hexagon picture, {at both the asymptotic and wrapping levels,} by confronting its predictions with  {available perturbative data. The focus will be on state-of-the-art correlators, involving two BPS operators and one non-BPS operator in the so-called $SL(2)$ sector, for which explicit results are culminating at three loops \cite{Eden, Emery}. There are several new effects, on the integrability side, showing up at this loop order precisely. It is the first time the dressing phase~\cite{BES}, which here enters as an ingredient in the hexagon form factor proposal~\cite{C123Paper}, contributes to the asymptotic part of the structure constant. It is also the first time on the wrapping side that some mirror channels open up. As already sketched in~\cite{C123Paper}, a single mirror particle passing through one of the edges adjacent to the non-BPS operator ($n_L=1$ or $n_R=1$ in figure \ref{hexagon}) first shows up at three loops. The same particle but in the edge opposed to the non-BPS operator ($n_B=1$ in figure \ref{hexagon}) shows up earlier, at two loops already. At three loops however we can access to the quantum corrected version of this process, and notably to the first effect of the mirror dressing phase. These are all the novel effects that will be studied here within the hexagon approach and confronted with perturbation theory. In all cases, as we shall see, a perfect match will be observed.}

\textit{Note added:} As we were writing up this work, we received the three loop analysis \cite{appeared} which overlaps substantially with some of our results. 

\begin{figure}[t]
\begin{center}
\includegraphics[scale=.6]{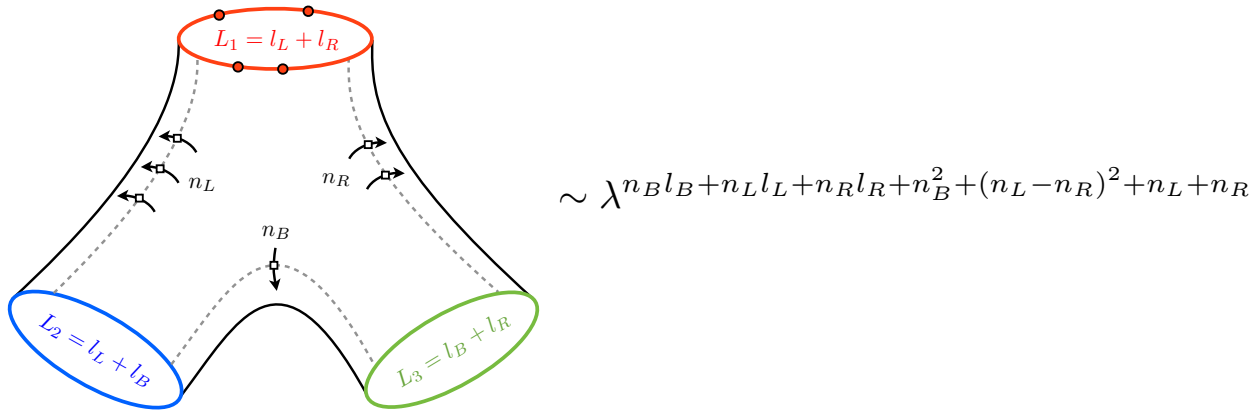}
\end{center}
\vspace{-7cm}
\caption{\normalsize  A pair of pants is cut into two hexagons \cite{C123Paper}. The excitations on the non-BPS operator (on the top)  can end up on either half and we should sum over those possibilities. 
Stitching the hexagons back into the pair of pants amounts to integrating over the various rapidities of the mirror particles at the dashed lines. A process with $n_L$ mirror excitations on the left dashed line, $n_R$ on the right and $n_B$ at the bottom shows up at $n_B l_B+n_L l_L+n_R l_R+n_B^2+(n_L-n_R)^2+n_L+n_R$ loops as indicated in the figure and explained in Appendix~\ref{MPI}. It is nice to note that the number of particles needed grows very slowly with perturbation theory. 
We see that up to three loop order, for instance, we can either have the vacuum in all dashed lines or a \textit{single} particle in a \textit{single} dashed line. The latter Luscher type corrections will only show up for very small bridges $l_{L}$, $l_R$ and $l_B$, related to the lengths of the three external operators as also indicated in the figure.
} \la{hexagon} \vspace{-.2cm} 
\end{figure}

\section{Data}
The comparison between theory (i.e. integrability) and experiment (i.e. direct perturbative computations) is one which involves compromise. From the integrability side, the simplest data to produce concern large operators, for which finite size corrections are suppressed. On the other hand, for perturbative computations, the smaller the external operator the  {simpler} are the underlying combinatorics. A nice set of correlators recently studied in \cite{Emery} provide an excellent middle ground. They are small enough to be efficiently computed in perturbation theory but large enough to allow us to have reasonable control over the various integrability finite size corrections. 

\begin{figure}[t]
\begin{center}
\includegraphics[scale=.6]{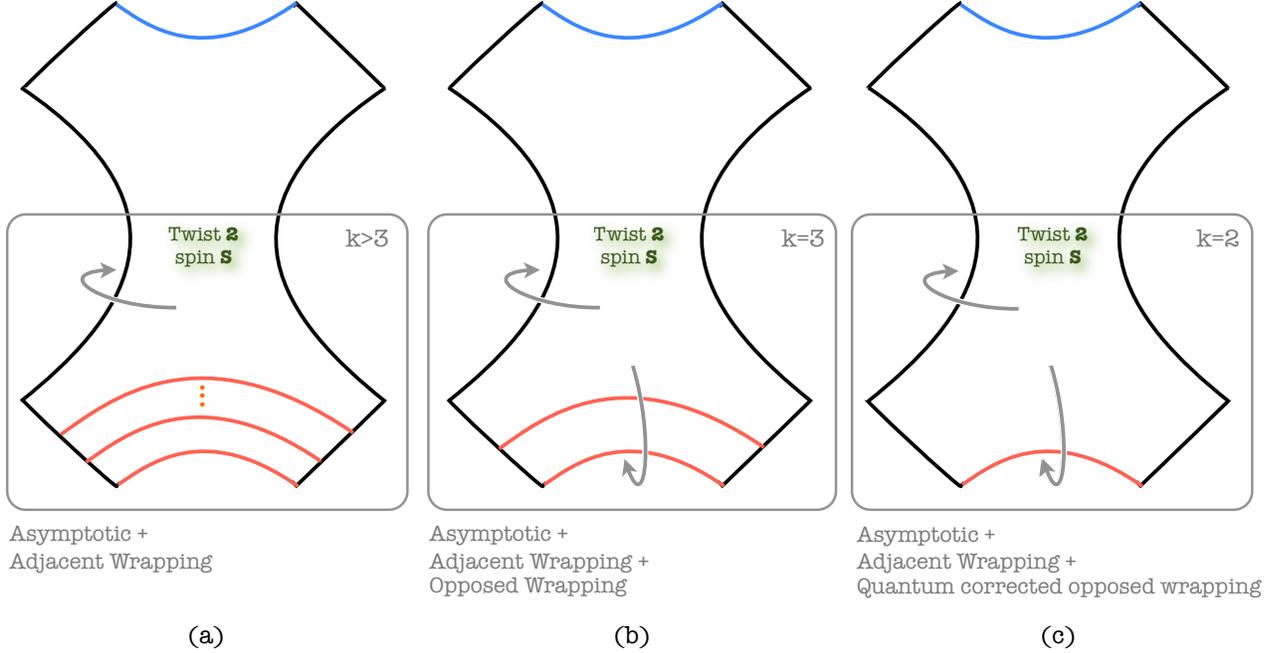}
\end{center}
\vspace{-4.5cm}
\caption{\normalsize 
{Tuning} $k$ in the four-point functions $\langle \textrm{Tr}(\bar{Z}X)\textrm{Tr}(\bar{Z}\bar{X})\textrm{Tr}(ZY^{k-1})\textrm{Tr}(Z\bar{Y}^{k-1})\rangle$  {studied in}~\cite{Emery}  {helps disentangling the} various finite size corrections  {to the \textit{three-loop} structure constants}. In all these cases we have $l_L=l_R=1$ such that, at three loops, we  {always} have to correct the asymptotic result by the adjacent mirror correction.  {The size of the} opposing wrapping correction  {however} depends  {interestingly} on $k$. 
For $k \ge 4$ we have $l_B \ge 3$ which completely suppresses this  {effect}. For $k=3$ the opposing bridge has length $l_B=2$ and the leading opposing wrapping correction is needed at three loops. Finally, the most complicated case from the integrability perspective is the rightmost one for $k=2$ corresponding to $l_B=1$ where we need to take into account quantum corrections to the opposed wrapping as well. From the perturbative size, complexity grows in the opposite direction, from right to left. 
} \la{FourPointFunction} \vspace{-.2cm} 
\end{figure}

\begin{figure}
\beq
\!\!\!\!\!\!\!\!\!\!\!\!\! \begin{array}{c|l}
S & \left.\(\frac{C_{123}^{\bullet\circ\circ}}{C_{123}^{\circ\circ\circ}}\)^2\right|_{l_B=3} \text{ for $l_L=l_R=1$ and spin $S$} \vspace{0.1cm}
\\ \hline
2 &\frac{1}{3}-4 g^2+56 g^4+g^6 \left(112 \zeta_3-160 \zeta _5-768\right)+\dots \color{white}{\Big(} \\
4 &\frac{1}{35}-\frac{205 g^2}{441}+\frac{73306 g^4}{9261}+g^6 \left(\frac{386 \zeta _3}{27}-\frac{400 \zeta _5}{21}-\frac{442765625}{3500658}\right)+\dots  \color{white}{\Big(}\\
6 &\frac{1}{462}-\frac{1106 g^2}{27225}+\frac{826643623 g^4}{1078110000}+g^6 \left(\frac{48286 \zeta _3}{37125}-\frac{56 \zeta _5}{33}-\frac{1183056555847}{88944075000}\right)+\dots  \color{white}{\Big(}\\
8 &\frac{1}{6435}-\frac{14380057 g^2}{4509004500}+\frac{2748342985341731 g^4}{42652702617525000}+g^6 \left(\frac{1039202363 \zeta _3}{9932422500}-\frac{6088 \zeta _5}{45045}-\frac{1270649655622342732745039}{1075922954067591630000000}\right)+\dots  \color{white}{\Big(}\\
10 & \!\frac{1}{92378}\!-\!\frac{3313402433 g^2}{13995964873800}\!+\!\frac{156422034186391633909 g^4}{31100584702491617040000}\!+\!g^6 \!\left(\!\frac{8295615163 \zeta _3}{1049947353000}-\frac{2684 \zeta _5}{264537}-\frac{7465848687069712820911408164847}{77747563297936585275804036000000}\right)\!+\!\dots  \color{white}{\Big(}
\end{array}  \nn
\eeq
\vspace{-.3cm}
\captionof{table}{Three loop structure constant $C_{44S}$ corresponding to a large bottom bridge $l_B=3$, see figure \ref{FourPointFunction}a. 
It is given by the asymptotic result plus the wrapping correction in the adjacent edge.
(In these tables we normalize the structure constants by the structure constants of three BPS scalar operators with the same R-charges.) This data is extracted from \cite{Emery} with $k=4$ (larger $k$'s yield the same) and matched against integrability in section~\ref{adjacent}.\la{table1}
}
\beq
 \begin{array}{c|l}
S & \left.\(\frac{C_{123}^{\bullet\circ\circ}}{C_{123}^{\circ\circ\circ}}\)^2\right|_{l_B=2}-\left.\(\frac{C_{123}^{\bullet\circ\circ}}{C_{123}^{\circ\circ\circ}}\)^2\right|_{l_B=3} \text{ for $l_L=l_R=1$ and spin $S$} \vspace{0.1cm}
\\ \hline
2 &80 g^6 \zeta _5+\dots \color{white}{\Big(} \\
4 &g^6 \left(\frac{4 \zeta _3}{3}+\frac{200 \zeta _5}{21}\right)+\dots  \color{white}{\Big(}\\
6 &g^6 \left(\frac{7 \zeta _3}{33}+\frac{28 \zeta _5}{33}-\frac{1}{180}\right)+\dots  \color{white}{\Big(}\\
8 &g^6 \left(\frac{3 \zeta _3}{130}+\frac{3044 \zeta _5}{45045}-\frac{79}{75600}\right)+\dots  \color{white}{\Big(}\\
10 & g^6 \left(\frac{781 \zeta _3}{366282}+\frac{1342 \zeta _5}{264537}-\frac{45071}{351630720}\right)+\dots \color{white}{\Big(}
\end{array}\nn
\eeq
\vspace{-.3cm}
\captionof{table}{
The structure constant $C_{33S}$ corresponds to a bottom bridge $l_B=2$. The difference between the structure constants $C_{33S} $ and $C_{44S}$ comes from the (leading) wrapping correction in the opposed channel, see figure \ref{FourPointFunction}b. This data is extracted from \cite{Emery} with $k=3$ and matched against integrability in section~\ref{oppSec}.  \la{table2}
}
\beq
 \begin{array}{c|l}
S &  \left.\(\frac{C_{123}^{\bullet\circ\circ}}{C_{123}^{\circ\circ\circ}}\)^2\right|_{l_B=1}-\left.\(\frac{C_{123}^{\bullet\circ\circ}}{C_{123}^{\circ\circ\circ}}\)^2\right|_{l_B=3} \text{ for $l_L=l_R=1$ and spin $S$} \vspace{0.1cm}
\\ \hline
2 &g^4\,24  \zeta _3-g^6 \left(240 \zeta _3+240 \zeta _5\right)+\dots \color{white}{\Big(} \\
4 &g^4 \left(\frac{20 \zeta _3}{7}+\frac{1}{3}\right)-g^6 \left(\frac{13318 \zeta _3}{441}+\frac{200 \zeta _5}{7}+\frac{655}{54}\right)+\dots  \color{white}{\Big(}\\
6 &g^4 \left(\frac{14 \zeta _3}{55}+\frac{199}{3960}\right)-g^6 \left(\frac{350413 \zeta _3}{136125}+\frac{28 \zeta _5}{11}+\frac{9984529}{4900500}\right)+\dots  \color{white}{\Big(}\\
8 &g^4 \left(\frac{1522 \zeta _3}{75075}+\frac{1721}{327600}\right)-g^6 \left(\frac{90199113551 \zeta _3}{473445472500}+\frac{3044 \zeta _5}{15015}+\frac{17141506511}{75125232000}\right)+\dots  \color{white}{\Big(}\\
10 &g^4 \left(\frac{671 \zeta _3}{440895}+\frac{578887}{1230707520}\right)- g^6 \left(\frac{3853245574541 \zeta _3}{293915262349800}+\frac{1342 \zeta _5}{88179}+\frac{846831496164217}{39443776036056000}\right)+\dots\color{white}{\Big(}
\end{array}\nn
\eeq
\vspace{-.3cm}
\captionof{table}{
The structure constant $C_{22S}$ corresponds to a small bottom bridge $l_B=1$. The difference between the structure constants $C_{22S} $ and $C_{44S}$ comes from the (leading and sub-leading) wrapping correction in the opposed channel, see figure \ref{FourPointFunction}c. This data is extracted from \cite{Eden} and matched against integrability in section~\ref{oppSec}.  \la{table3}
}
\end{figure}

More precisely, in \cite{Emery} four point functions $\langle \mathcal{O}_2\mathcal{O}_2\mathcal{O}_k\mathcal{O}_k\rangle $ -- involving two small BPS operators of size $2$ (commonly referred to as $20'$ operators) and two BPS operators of size $k$ -- were studied up to three loops. (The case $k=2$ was known before \cite{EdenHidden,Drummond4pt}.)
From the OPE decomposition of these  {ones} we can read  {off} the product of structure constants $C_{22S}\times C_{kkS}$ where $S$ denotes the spin of the lowest twist operator being exchanged (which are twist two operators in this case), see figure \ref{FourPointFunction}.

In the end, from this three-loop data, we produce tables \ref{table1}--\ref{table3}. The main goal of this paper is to reproduce these tables using the integrability approach by taking into account the various new physical effects which kick in at this loop order.

\begin{figure}[t]
\begin{center}
\includegraphics[scale=.38]{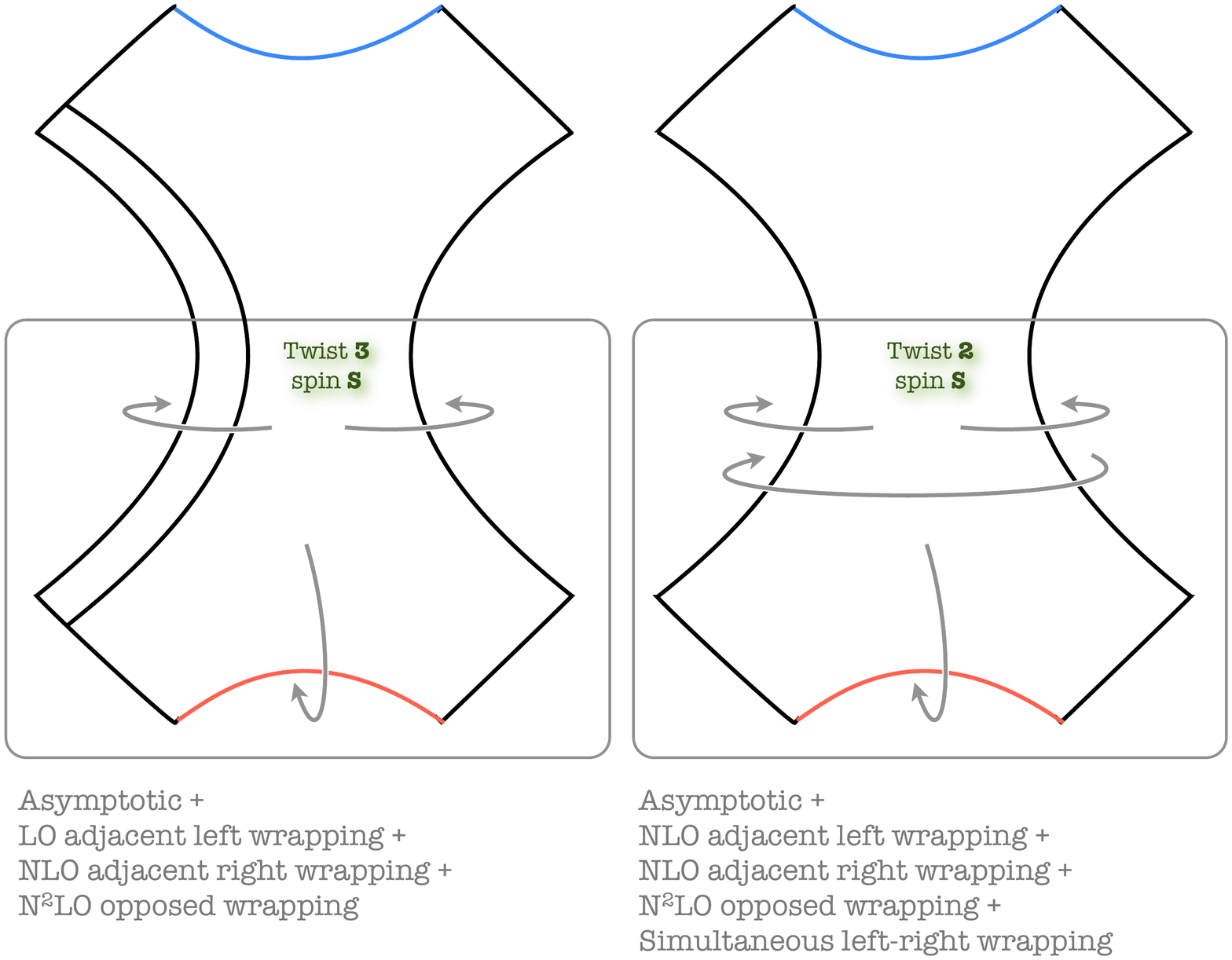}
\end{center}
\vspace{-0.2cm}
\caption{\normalsize 
At \textit{four loops}, from a four point function $\langle \textrm{Tr}(\bar{Z}X)\textrm{Tr}(\bar{Z}\bar{X})\textrm{Tr}(ZY^{k-1})\textrm{Tr}(Z\bar{Y}^{k-1})\rangle$ with $k=2$ (on the right) we could test a very interesting new effect: the simultaneous wrapping effect. It is quite non-trivial to dig it out of a background of various other contributions. As a warm up exercise, it would be very convenient to also have at our disposal the case with $k=3$ (on the left) with which we could first test all other complicated but presumably well understood effects before attacking the most interesting $k=2$ case. See conclusions for a more detailed discussion on the importance of these checks. 
} \la{wishList} \vspace{-.2cm} 
\end{figure}

As discussed in more detail in the conclusions there are various fascinating effects to probe at even higher loops. We look forward, in particular, to having four loop data to analyze. In this regard, the simplest case to compute from perturbation theory is probably the $k=2$ case corresponding to four external operators in the stress tensor multiplet. After all, for these  {ones} the result is already known \cite{EdenHidden} in terms of a few unknown integrals which one would have to evaluate (or at least to work out  {in the OPE limit}). From the integrability point of view having a few other examples with larger operators would be useful as well. For example, the case $k=3$ would allow us to isolate and check all effects other than simultaneous wrapping, see figure \ref{wishList}. It would be simpler to sharpen the integrability machinery with that case first. Of course, with even larger external operators we could  {disentangle further the} various finite size corrections and increase the complexity of the integrability computation in an even more controllable fashion. The more (data) the merrier.

\section{Integrability} \la{IntegrabilitySec}

The basic idea of the hexagon program is to cut the three-point function into two overlapping hexagons along three mirror edges. The recipe for sewing them back together is to perform a sum over complete basis of states for each mirror channel. As a result, the three-point function is expressed as an infinite series over mirror particles, each term of which comes with the {\it measure}, which is the cost of producing particles, and the factor needed for propagating particles from one hexagon to the other through the so-called {\it bridges} of size $\ell_{ij}$. The leading term in the series corresponds to taking the mirror vacuum in all the channels and is called the {\it asymptotic three-point function}. It dominates both in the large bridge limit, for the obvious reason that large distance suppresses the propagation of gapped excitations, and in the weak coupling limit where the production and the propagation are typically suppressed. 

To make this program run at higher loops, one must have a good handle on complicated hexagon processes involving many excitations exchanged in the three mirror channels. Although intractable at first glance, they in fact have a simple nice structure: Once we perform the summation over the flavour indices of mirror particles, the contribution from each mirror excitation becomes proportional to the $\mathfrak{psu}(2|2)$ transfer matrix (see Appendix \ref{MPI} and figure \ref{YB2} for more detailed explanation). This is due to the Yang-Baxter relation and is a direct consequence of the matrix structure of our ansatz, which is essentially equivalent to the $\mathfrak{psu}(2|2)$ S-matrix. The constant of proportionality can then be read off by studying the cases where mirror particles are longitudinal derivatives $D$. As explained in Appendix \ref{MPI}, the end result is remarkably simple, and the multi-particle integrand for fundamental mirror particles takes the following form\footnote{Here a function with sets as arguments denote a product of such functions with the elements of the sets as arguments. For explicit definition, see (\ref{oftenusednotation}).}:
\beqa
&& 
\!\!\!\!\!\!\!\!\! \texttt{Integrand}=\mu (\bottom{\wB^{\gamma}}) e^{-E (\bottom{\wB})l_B}T(\bottom{\wB^{\gamma}})h^{\neq} (\bottom{\wB^{\gamma}},\bottom{\wB^{\gamma}})    h({\bf u},\bottom{\wB^{-3\gamma}})\, \times \la{mainformula} \\
&&\!\!\!\!\!\!\!\!\!\! \qquad  \mu (\lefta{\wL^{\gamma}})e^{-E (\lefta{\wL}) l_L} T(\lefta{\wL^{-\gamma}}) h^{\neq} (\lefta{\wL^{\gamma}},\lefta{\wL^{\gamma}})   \times \mu (\righta{\wR^{\gamma}}) e^{-E (\righta{\wR}) l_R} T(\righta{\wR^{-\gamma}})h^{\neq} (\righta{\wR^{\gamma}},\righta{\wR^{\gamma}}) \,\times \nn\\
&&\!\!\!\!\!\!\!\!\!\! \qquad  h(\lefta{\wL^{-\gamma}},\righta{\wR^{-5\gamma}})h(\righta{\wR^{-\gamma}},\lefta{\wL^{-5\gamma}}) \! \sum_{\alpha\cup \bar{\alpha}={\bf u}}\! (-1)^{|\bar{\alpha}|}e^{ip_{\bar{\alpha}}l_R}\frac{h(\alpha,\lefta{\wL^{-5\gamma}})h(\alpha,\righta{\wR^{-\gamma}})h(\bar{\alpha},\lefta{\wL^{-\gamma}})h(\bar{\alpha},\righta{\wR^{-5\gamma}})}{h(\alpha,\bar{\alpha})}\,. \nn
\eeqa
Here ${\bf u}$ is a set of rapidities for physical excitations and $\wB$, $\wL$ and $\wR$ denote {the sets of mirror} rapidities for the bottom, the left adjacent and the right adjacent edges respectively. For bound states, we just need to substitute $h$, $\mu$, $T$ and $E$ in (\ref{mainformula}) with their bound-state counterparts, which are given in Appendix \ref{FHandWCE}. 

{As explained in more detail in appendix \ref{MPI}, taking into account the scaling with the coupling of the various terms in the integrand, the estimate in figure \ref{hexagon} follows straightforwardly. As mentioned above, up to three loops, we can restrict the integrand to at most a single particle in either mirror edge. 
It what follows we will study these processes at three loops and see that they perfectly match with the perturbative data presented in the previous section. }

\subsection{Asymptotic Result}
At leading order in the large distance expansion, only the vacuum states in the mirror channels contribute. We can thus 
set $n_B=n_L=n_R=0$ in (\ref{mainformula}) and reduce it to the asymptotic structure constants \cite{C123Paper},
\beq
\left. \frac{C_{123}^{\bullet \circ \circ}}{C_{123}^{\circ \circ \circ}}\right|_\text{asymptotic} = \underbrace{\sqrt{ \frac{\prod\limits_{i} \mu_i  \prod\limits_{i\neq j}h_{ij}}{
\underset{1\le i,j \le S}{\det}
\, \partial_{u_i} \Big(p_{j}L+\frac{1}{i}\sum\limits_{k \neq j} \log S_{jk}\Big)}} }_{\equiv \,\texttt{Gaudin}}\times
\underbrace{\sum_{\alpha\cup \bar{\alpha}={\bf u}} (-1)^{|\bar{\alpha}|} \prod_{j\in \bar{\alpha}}  e^{ip_{j}\ell} \prod_{i\in \alpha, j\in \bar{\alpha}} \frac{1}{h_{ij}} }_{\equiv\, \mathcal{A}_\text{asymptotic}} \, , \la{rank1}
\eeq
where $L=L_1$ is the twist of non-BPS operator (of spin $S$) and $\ell=l_R$ is the length of one of the adjacent bridges\footnote{It can be either $l_R$ or $l_L=L-l_R$; Bethe equations together with $S_{ij}=h_{ij}/h_{ji}$ ensure that the result is the same.}. The determinant factor inside the square root is the famous Gaudin norm. The most interesting factor in the structure constant is the last factor. It is given by a sum over the distributions of the Bethe roots into two partitions which arises from cutting  {the} pair of pants into two, see figure \ref{hexagon}. We are using the short-hand notation $\mu_k=\mu(u_k)$, $S_{ij}=S(u_i,u_j)$, $h_{ij}=h(u_i,u_j)$ etc. Explicit expressions for the measure $\mu$ and hexagon transitions $h$ can be found in \cite{C123Paper} and are summarized in Appendix \ref{FHandWCE} for convenience.

Here we are interested in the case of twist $L=2$ and adjacent bridges $\ell=l_R=l_L=1$. Solving Bethe equations is particularly simple for twist $2$. There is a single solution for each spin $S$ which is best encoded in the so-called Baxter polynomial $Q(u) \equiv \prod_{j=1}^S (u-u_j)$. To find this polynomial to high order in perturbation theory we can solve Bethe equations in Mathematica with very high precision and rationalize the final result. For spin $S=4$, for instance, we find
\beqa
Q(u)&=&\(u^4-\frac{13
   }{14}\,u^2+\frac{27}{560}\)+g^2
   \left(\frac{60}{49}-\frac{384 }{49}\,u^2\right)
+g^4 \left(\frac{7370
   }{1029}\,u^2+\frac{7990}{1029}\right) \nn \\
   &+& g^6 \left(u^2 \left(-\frac{200 \zeta_3}{7}-\frac{335225}{7203}\right)+\frac{50 \zeta_3}{7}+\frac{21325}{28812}\right)+O(g^8)
\eeqa
Note the appearance of $\zeta_3$ in the three loop correction to the Bethe roots. It comes from the BES dressing phase which first shows up at this loop order. (This effect shows up for three-point functions at three loops but affects the spectrum at four loops only since the magnon dispersion relation is itself of order $g^2$.) The procedure is then straightforward. For each spin $S$ we solve Bethe equations and plug the Bethe roots in (\ref{rank1}). (At lower loop orders, this is spelt out in great detail in \cite{C123SL2} and \cite{C123Paper}.)  In this way we generate the following table:
\beq
\!\!\!\!\!\!\!\! \begin{array}{c|l}
S & \left.\(\frac{C_{123}^{\bullet\circ\circ}}{C_{123}^{\circ\circ\circ}}\)^2\right|_\text{asymptotic} \text{ for $l_L=l_R=1$ and spin $S$} \vspace{0.1cm}
\\ \hline
2 &\frac{1}{3}-4 g^2+56 g^4+g^6 \left(
-804 + 16 \,\zeta_3\right)+\dots \color{white}{\Big(} \\
4 &\frac{1}{35}-\frac{205 g^2}{441}+\frac{73306 g^4}{9261}+g^6 \left(
\frac{134 \zeta_3}{63}-\frac{3670467025}{28005264}
\right)+\dots  \color{white}{\Big(}\\
6 &\frac{1}{462}-\frac{1106 g^2}{27225}+\frac{826643623 g^4}{1078110000}+g^6 \left(
\frac{1484 \zeta_3}{7425}-\frac{4879310394853}{355776300000}\right)+\dots  \color{white}{\Big(}\\
8 &\frac{1}{6435}-\frac{14380057 g^2}{4509004500}+\frac{2748342985341731 g^4}{42652702617525000}+g^6 \left( \frac{4665511 \zeta_3}{283783500}-\frac{10449826286558318778958087}{860738363254073304
   0000000} \right)+\dots  \color{white}{\Big(}\\
10 & \!\frac{1}{92378}\!-\!\frac{3313402433 g^2}{13995964873800}\!+\!\frac{156422034186391633909 g^4}{31100584702491617040000}\!+\!g^6 \!\left(\frac{21027743 \zeta_3}{16665831000}-\frac{61273849341907187613352885884203}{6219805063
   83492682206432288000000}\right)\!+\!\dots  \color{white}{\Big(}
\end{array}   \nn
\eeq
{\captionof{table}{
Integrability predictions for the asymptotic result for twist two operators.  \la{table1Int} \vspace{.4cm} } }

We see that up to order $g^4$, i.e. two loops, this table is identical to table \ref{table1}. This is expected and was already pointed out in \cite{C123Paper}. After all, table \ref{table1} concerns three-point functions with a large bottom bridge $l_B>2$ for which finite size corrections should kick in at three loops (for the adjacent channels and at even higher loops for the opposing channel). In the next section \ref{adjacent}, we will see that once these adjacent wrapping corrections are added to the asymptotic result, table~\ref{table1} is perfectly reproduced. In section \ref{oppSec} we study the opposing wrapping which becomes relevant at three loops provided one decreases the length of the bottom bridge to $l_B \le 2$. In this way we will reproduce the remaining perturbative data quoted in tables \ref{table2} and \ref{table3}.

\subsection{Adjacent Wrapping} \la{adjacent}
We now move to the first adjacent wrapping correction corresponding to putting a single mirror particle in either of the adjacent channels, i.e. for $n_L=1$ or $n_R=1$ with all other occupation numbers set to zero. This new wrapping effect starts at three loops for adjacent bridges $l_L=l_R=1$ as highlighted in figure \ref{hexagon} so it is quite interesting to test it against fresh perturbative data. From (\ref{mainformula}) we have
\beqa
 \left.\(\frac{C_{123}^{\bullet\circ\circ}}{C_{123}^{\circ\circ\circ}}\)^2\right|_\text{adjacent}  &=& \texttt{Gaudin} \times  \sum_{a=1}^\infty \int  \! dv\, \frac{T_a(v^{-\gamma})\mu_a(v^\gamma) }{(y^{[a]}y^{[-a]})^\ell \prod\limits_{j} h_a(v^{-\gamma},u_j)}\! \times\nn\\ &&\qquad \times
\underbrace{ \sum_{\alpha\cup \bar{\alpha}={\bf u}} (-1)^{|\alpha|} \prod_{j\in \bar{\alpha}} (e^{ip_{j}\ell}h_a(u_j,v^{-\gamma})h_a(v^{-\gamma},u_j))\!\! \prod_{i\in \alpha, j\in \bar{\alpha}}  \frac{1}{h_{ij}}  }_{\equiv \,\mathcal{A}_\text{adjacent}(v)} \, , \la{adjacent}
\eeqa
where $\ell=l_L=l_R$ ($=1$ for the twist two case of interest for this paper). We should now expand these expressions in perturbation theory, perform the $v$ integration and sum over the bound-state index $a$. 

Amusingly, the last two of these steps turn out to be trivial as we now explain.  The last line in (\ref{adjacent}) resembles the sum over partitions which we encountered in the asymptotic regime in the last subsection with an effective momentum term $e^{i p(u) \ell} \to e^{i p(u) \ell}  h_a(u,v^{-\gamma})h_a(v^{-\gamma},u)$. With this effective momentum, the $\mathcal{A}_\text{adjacent}$ factor actually yields a very familiar object: another copy of the mirror transfer matrix $T_a(v^{-\gamma})$ in the $a$-th anti-symmetric representation! That is, at weak coupling, we find that when $\ell=1$
\beq\la{Aadjacent}
\mathcal{A}_\text{adjacent}(v)  / \mathcal{A}_\text{asymptotic} = \frac{(-1)^{a}v^{[a]}v^{[-a]}Q(i/2)}{-g^2 \gamma Q^{[1+a]}} T_{a}(v^{-\gamma}) +O(g^2) \,.
\eeq
The proof of this interesting identity will be given in Appendix \ref{Tfromsum}.
Taking into account the weak coupling expressions for the transfer matrix, measures and for the dynamical parts (presented in appendices \ref{Tap} and \ref{FHandWCE}) we have therefore 
\beq
\!\!\!  \left.\(\frac{C_{123}^{\bullet\circ\circ}}{C_{123}^{\circ\circ\circ}}\)^2\right|_\text{adjacent}\!\!\!  =  \left.\(\frac{C_{123}^{\bullet\circ\circ}}{C_{123}^{\circ\circ\circ}}\)^2\right|_\text{asymptotic} \!\!    \times \frac{g^6}{\gamma} \times \left[ \sum_{a=1}^\infty \int \! dv  \frac{- \frac{\gamma^2}{2}  \Big(  Q(\tfrac{i}{2}) \sum\limits_{k=-(a-1)/2}^{(a-1)/2} \displaystyle \frac{Q^{[2k]}}{v^{[2k+1]}v^{[2k-1]}} \Big)^2 }{(v^{[a]}v^{[-a]}
)^2 Q^{[-a+1]}Q^{[a-1]}Q^{[-a-1]}Q^{[a+1]}} \right] \nn
\eeq
Now, beautifully, the object in the square brackets is the \textit{exact same} integral and sum which yields the wrapping correction to the energy of multi-particle states \cite{JanikKonishi} as computed by Bajnok, Janik and Lukowski in \cite{Janik}. In this work, the authors also evaluated this correction for twist~$2$ operators of arbitrary spin~$S$. Therefore, to get the adjacent finite size correction to the structure constants, we simply need to divide their result by the one loop anomalous dimension $\gamma=8S_1$. (Curiously, this very same object is appearing at four loops for the spectrum corrections and at three loops for the structure constants; the division by $\gamma$ nicely ensures that the transcendentality counting works.)
In this way we can immediately generate the following table:

\beq
\!\!\!\!\!\!\!\! \begin{array}{c|l}
S & \left.2\(\frac{C_{123}^{\bullet\circ\circ}}{C_{123}^{\circ\circ\circ}}\)^2\right|_\text{adjacent} \text{ for $l_L=l_R=1$ and spin $S$} \vspace{0.1cm}
\\ \hline
2 & g^6 \left( 36 + 96 \zeta_3 - 160 \zeta_5\right)+\dots \color{white}{\Big(} \\
4 &g^6 \left( \frac{2300 \zeta_3}{189}-\frac{400  \zeta_5}{21}+\frac{41575}{9072}\right)+\dots \color{white}{\Big(} \\
6 &g^6 \left( \frac{13622 \zeta _3}{12375}-\frac{56 \zeta
   _5}{33}+\frac{7367101}{17820000}\right)+\dots \color{white}{\Big(} \\
8 &g^6 \left( \frac{145984913 \zeta _3}{1655403750}-\frac{6088 \zeta
   _5}{45045}+\frac{8828613403153}{266983516800000}\right)+\dots \color{white}{\Big(} \\
10 &g^6 \left( \frac{3485433677 \zeta _3}{524973676500}-\frac{2684 \zeta
   _5}{264537}+\frac{47383910636511053}{19050244772832000000}\right)+\dots \color{white}{\Big(} 
\end{array}  \nn
\eeq
{\captionof{table}{
Integrability predictions for \textit{twice} the adjacent wrapping for twist two operators. The factor of two is convenient since  {we} have a left and a right adjacent contribution which yield the same result.   \la{table2Int} \vspace{.4cm}
}} 

It is now a very pleasurable task to add up the two integrability generated tables~\ref{table1Int} and~\ref{table2Int} and observe that they perfectly reproduce the OPE data in table \ref{table1}! This is another important check of the hexagon proposal; it is the first non-trivial check of the so-called adjacent wrapping corrections.

\subsection{Opposing Wrapping}  \la{oppSec}
We next study the contribution from the opposing channel. Up to {five} loops, we have only single-particle wrapping corrections and the integrand can be obtained by setting~$n_B=1$ and $n_L=n_R=0$ in (\ref{mainformula}):
\beq\la{opposing-all-loop}
\begin{aligned}
\left.\(\frac{C_{123}^{\bullet\circ\circ}}{C_{123}^{\circ\circ\circ}}\)^2\right|_\text{opposing}  &=& \texttt{Gaudin} \times \mathcal{A}_{\rm asymptotic}\times  \sum_{a=1}^\infty \int  \! dv\, \frac{T_a(v^{\gamma})\mu_a(v^\gamma) }{(y^{[a]}y^{[-a]})^\ell \prod\limits_{j} h_a(v^{\gamma},u_j)}
\end{aligned}
\eeq
The opposing wrapping already shows up at two loop for $l_B=1$, as studied in \cite{C123Paper}, and it appears for both $l_B=1$ and $l_B=2$ at three loop. For $l_B=2$, we just need to keep the leading term in the expansion of (\ref{opposing-all-loop}) and integrate. This is essentially the same as the two loop computation for $l_B=1$ and we can use the same methodology explained in \cite{C123Paper}. On the other hand, for $l_B=1$, we need to expand it one-loop further. This results in a more complicated integrand, for which analytic integration is harder to perform. To overcome this difficulty, we develop a simple trick which is explained in detail in Appendix \ref{HPL}: The basic idea is to replace the Baxter polynomial $Q(v)$ appearing in the integrand with the ``plane-wave'' form, $e^{iu t}$. The integral with this plane-wave expression is more convergent than the original one and we can simply compute it by taking the residues. Once it is computed, we can retrieve the results for any Baxter polynomials by simply applying the differential operator $Q(-i\partial_{t})$ to the final answer. 

With this new trick, it is now straightforward to generate results for any spin and they are summarized in the following tables:
\beq
\!\!\!\!\!\!\!\! \begin{array}{c|l}
S & \left. \(\frac{C_{123}^{\bullet\circ\circ}}{C_{123}^{\circ\circ\circ}}\)^2\right|_\text{opposing} \text{ for $l_B=2$, $l_L=l_R=1$ and spin $S$} \vspace{0.1cm}
\\ \hline
2 &80 g^6 \zeta _5+\dots \color{white}{\Big(} \\
4 &g^6 \left(\frac{4 \zeta _3}{3}+\frac{200 \zeta _5}{21}\right)+\dots  \color{white}{\Big(}\\
6 &g^6 \left(\frac{7 \zeta _3}{33}+\frac{28 \zeta _5}{33}-\frac{1}{180}\right)+\dots  \color{white}{\Big(}\\
8 &g^6 \left(\frac{3 \zeta _3}{130}+\frac{3044 \zeta _5}{45045}-\frac{79}{75600}\right)+\dots  \color{white}{\Big(}\\
10 & g^6 \left(\frac{781 \zeta _3}{366282}+\frac{1342 \zeta _5}{264537}-\frac{45071}{351630720}\right)+\dots \color{white}{\Big(}
\end{array}  \nn
\eeq
{\captionof{table}{
Integrability predictions for the opposing wrapping for an opposing bridge of size $l_B=2$. It beautifully matches with perturbative data in table \ref{table2}   \vspace{.4cm}
\la{table6}}} 
\vspace{-1cm}
\beq
\!\!\!\!\!\!\!\! \begin{array}{c|l}
S & \left. \(\frac{C_{123}^{\bullet\circ\circ}}{C_{123}^{\circ\circ\circ}}\)^2\right|_\text{opposing} \text{ for $l_B=1$, $l_L=l_R=1$ and spin $S$} \vspace{0.1cm}
\\ \hline
2 &g^4\,24  \zeta _3-g^6 \left(240 \zeta _3+240 \zeta _5\right)+\dots \color{white}{\Big(} \\
4 &g^4 \left(\frac{20 \zeta _3}{7}+\frac{1}{3}\right)-g^6 \left(\frac{13318 \zeta _3}{441}+\frac{200 \zeta _5}{7}+\frac{655}{54}\right)+\dots  \color{white}{\Big(}\\
6 &g^4 \left(\frac{14 \zeta _3}{55}+\frac{199}{3960}\right)-g^6 \left(\frac{350413 \zeta _3}{136125}+\frac{28 \zeta _5}{11}+\frac{9984529}{4900500}\right)+\dots  \color{white}{\Big(}\\
8 &g^4 \left(\frac{1522 \zeta _3}{75075}+\frac{1721}{327600}\right)-g^6 \left(\frac{90199113551 \zeta _3}{473445472500}+\frac{3044 \zeta _5}{15015}+\frac{17141506511}{75125232000}\right)+\dots  \color{white}{\Big(}\\
10 &g^4 \left(\frac{671 \zeta _3}{440895}+\frac{578887}{1230707520}\right)- g^6 \left(\frac{3853245574541 \zeta _3}{293915262349800}+\frac{1342 \zeta _5}{88179}+\frac{846831496164217}{39443776036056000}\right)+\dots\color{white}
\end{array}  \nn
\eeq
{\captionof{table}{
Integrability predictions for the opposing wrapping for an opposing bridge of size $l_B=1$.   It perfectly matches with perturbative data in table \ref{table3} \vspace{.4cm}
\la{table7}}} 

Having computed all the relevant wrapping corrections at three loop, we can now compare them with the   perturbative data. For this purpose, it is convenient to subtract the asymptotic structure constant and the adjacent wrappings from the perturbative data. In the case at hand, this can be achieved simply by subtracting the perturbative result for $l_B=3$ from the relevant data since the structure constants with $l_B=3$ do not receive the opposing wrapping correction at three loop. This is precisely what is done in table \ref{table2} and table \ref{table3}. The integrability predictions just produced, in tables \ref{table6} and \ref{table7}, beautifully match with the perturbative data. 

\section{Conclusion}\la{conclusion}

In this paper, we  {successfully compared} the hexagon program against perturbative data up to three loops  {and obtained the general expression for the hexagon integrand with arbitrarily many particles in each mirror channel. The form of the latter clearly indicates several important milestones, at even higher loop orders, which call for more perturbative data and stand as a challenge for (or might lead to some amendments to) the hexagon program. See figure \ref{examples} for a road map. }

The next   {important test} {in line, and perhaps the most critical one, awaits us already at the next loop order, that is at four loops.} At four loops we can have one excitation in each of the two adjacent edges to the non-BPS operator,  {as shown in figure \ref{examples}}. Pictorially this is  {the same} sort of  {drawing as} for the usual wrapping corrections arising in the spectrum,  {describing a single mirror particle crossing once each of the two dashed lines and thus winding once around the non-BPS operator.} It is not a coincidence that this sort of \textit{full} wrapping effect first shows up at the same loop order for the spectrum and for the 3-point function,  {since the SUSY cancellations delaying it to four loops is the same in both cases.} 

 {The hard task with this type of effect (involving at least one particle in each dashed line) is that it leads to a (double) pole in the hexagon integrand. This one comes about} when the two mirror rapidities we are integrating over collide  {and is manifest in}~(\ref{mainformula}) where it arises from $1/h(w^{\gamma}_R,w^{\gamma}_L)h(w^{\gamma}_L,w^{\gamma}_R)$ and from the fact that $h(u^{\gamma},v^{\gamma})$ vanishes for equal rapidities.  {An optimistic point of view would be} that  {this} pole  {just calls for a} prescription, such as principal part integration, for instance.\footnote{In the so-called pentagon OPE approach \cite{POPE} for scattering amplitudes, there is a similar decoupling pole for the pentagon which shows up in the heptagon  {and higher n-gons integrands}. In that case  {however there is a well-understood $i\epsilon$ prescription for integrating it}.}  {It might also well be the tip of an iceberg}
\begin{figure}[t]
\begin{center}
\includegraphics[scale=.25]{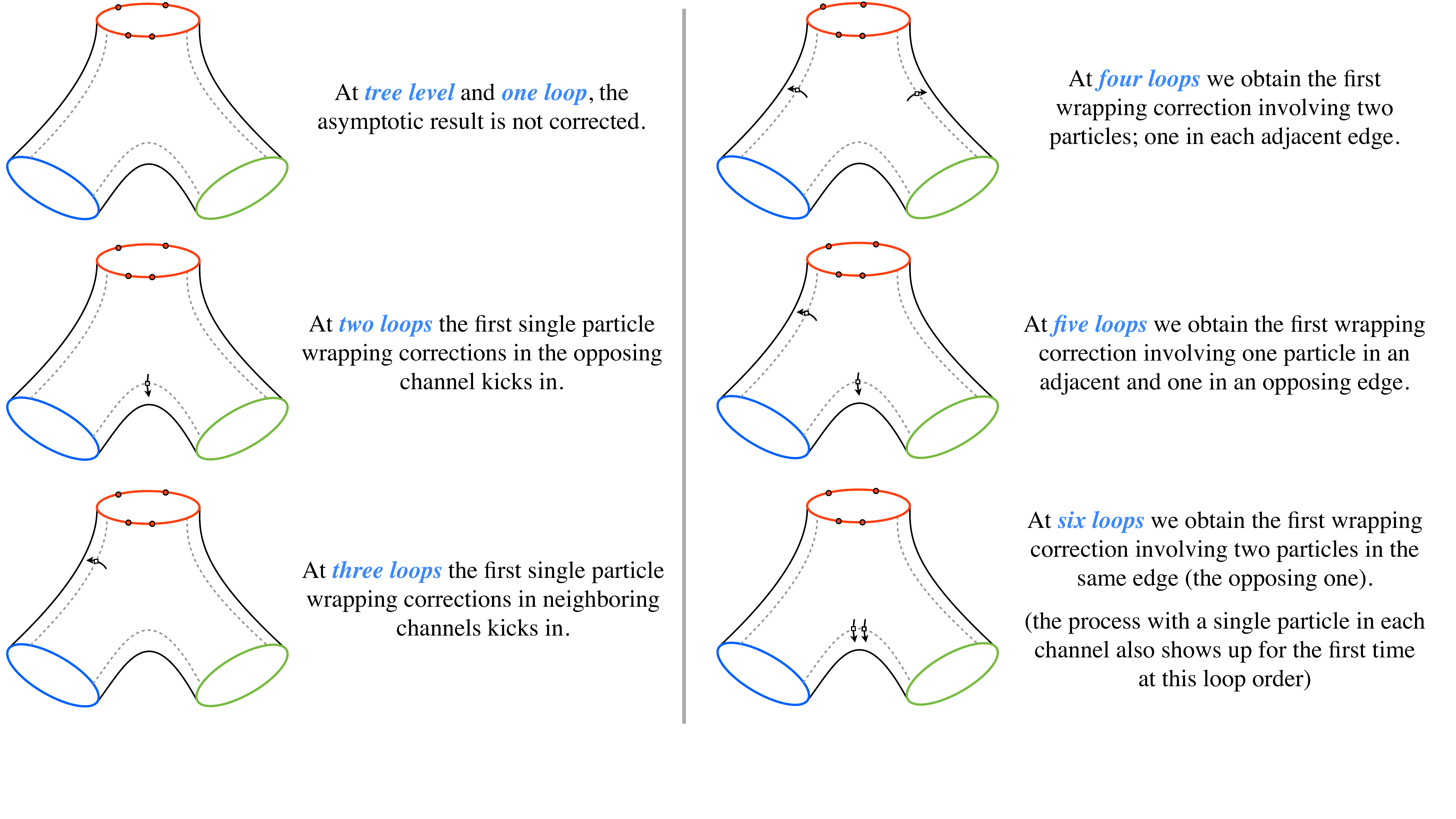}
\end{center}
\vspace{-1cm}
\caption{\normalsize  Various new physical effects await us at the next few loop orders. The estimates here hold for the smallest possible external operators (all with length two). For bigger operators these effects are delayed as summarized in figure \ref{hexagon}.
} \la{examples} \vspace{-.2cm} 
\end{figure}
and the symptom of something yet to be understood. Four loops is, for instance, the onset of corrections to the Bethe wave function as well, which might challenge the splitting procedure as understood so far and produce corrections to the asymptotic Gaudin norm entering our expression. Can it be that the smoothing of the singularity just discussed requires taking this into account? This is definitely plausible. After all, the  pole at equal  {momenta results from} a decoupling limit  {with a residue coming from} a single mirror particle which encircles the non-BPS operator asymptotically close to it.%
\footnote{ {As such the pole is an IR effect which comes about because the mirror space is taken to have infinite volume}.  {This is reminiscent of the IR ambiguities that plague the form factor approach to finite temperature correlators and must also be handled carefully~\cite{FFFV1,FFFV2}.}}
 {But is it not} exactly how a putative wrapping correction to the norm would look like? Clearly, having four loops gauge theory data at hand would be  {a fantastic help for settling all that.} It would allow  {one} to experiment  {the various} options and hopefully figure out which one holds best.  

 {At higher loops, the interplay between perturbative and integrability methods could also help better} understand the space of functions appearing in the four-point correlators  {and perhaps design} a program akin to the hexagon  function program \cite{LanceEtAl1} which  {appears so} powerful  {for scattering amplitudes,} in conjunction with the pentagon OPE \cite{POPE}. Perhaps a less ambitious program would be to develop such understanding close to the so-called light-cone OPE limit since this is the limit required to extract data about the leading twist operators flowing in the conformal partial wave decomposition.

Looking ahead, once all stitching subtleties are properly understood we can start cutting and gluing any strings that move. Four-point functions, for instance, would be a very natural next frontier. 

\section*{Acknowledgements} 
We are grateful to E.~Sokatchev for valuable discussions and for sharing with us extremely valuable unpublished three-loop data. 
Research at the Perimeter Institute is supported in part by the Government of Canada through NSERC and by the Province of Ontario through MRI. 

\appendix
\section{Multi-particle Integrand}\label{MPI}

Here we derive an integrand for general multi-particle wrapping corrections for a class of correlators studied in the main text. For simplicity, below we focus on fundamental mirror particles, but the results can be extended easily to bound states.
 \begin{figure}[t]
\begin{center}
\includegraphics[scale=.4]{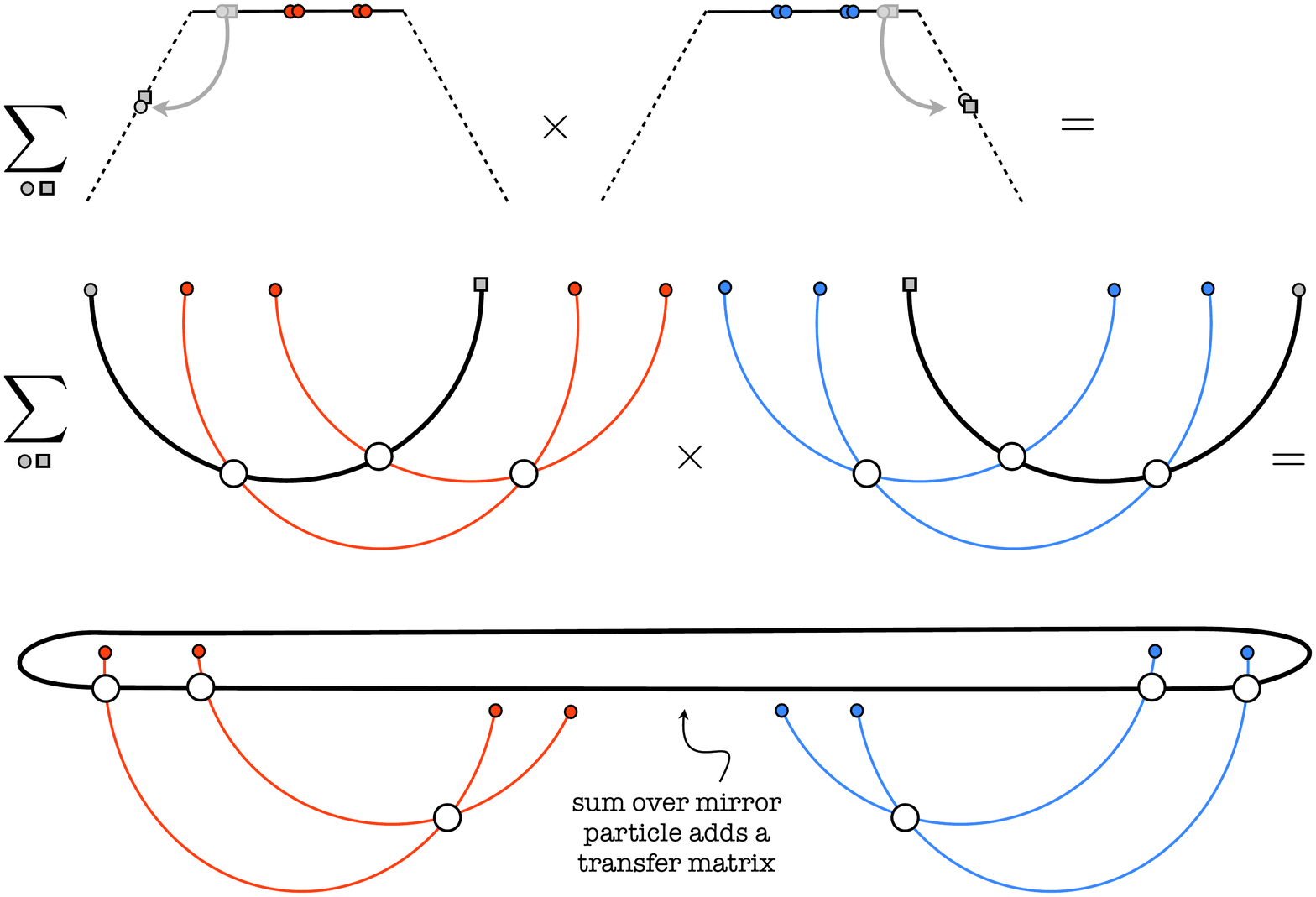}
\end{center}
\vspace{-1cm}
\caption{\normalsize Matrix part for a single mirror excitation. After summing over flavour indices, the wrapping correction becomes proportional to the $\mathfrak{psu}(2|2)$ transfer matrix.} \la{YB1} \vspace{-.2cm} 
\end{figure}
Before delving into general multi-particle cases, we shall briefly recall the result in \cite{C123Paper} for a single-particle wrapping on the bottom edge. Such a correction can be computed by putting a mirror particle on the first hexagon and its anti-particle on the second hexagon. A crucial observation made in \cite{C123Paper} is that, after summing over flavours of the particle ($D$, $\bar{D}$, $Y$ and $\bar{Y}$), the result becomes proportional to a transfer matrix $T(u^{-\gamma})$. This is a direct consequence of the matrix structure of the hexagon form factor and can be understood pictorially as in figure \ref{YB1}. The constant of proportionality is then determined by evaluating the configuration with $D$ in the first hexagon and $\bar{D}$ in the second hexagon. For the bottom edge, it is given simply by a product of $h$'s and the asymptotic structure constant (\ref{rank1})\footnote{The expression (\ref{single-integrand}) looks slightly different from the one written in \cite{C123Paper}. However, it can be recast into the same form by using $h(u,v)=1/h(v^{4\gamma},u)$}:
\beq\label{single-integrand}
h({\bf u},\wB^{-3\gamma})\times \left[{\tt Gaudin} \times \sum_{\alpha \cup \bar{\alpha}={\bf u}}(-1)^{|\bar{\alpha}|}\prod_{j\in \alpha}e^{ip_j l_{R}}\frac{1}{h(\alpha,\bar{\alpha})}\right]\,.
    \eeq
    Here and below, we are using simplified notations such as
    \beq\la{oftenusednotation}
    \begin{aligned}
    h({\bf u},{\bf v})\equiv &\prod_{u_i\in {\bf u},v_j\in{\bf v}} h(u_i,v_j)\,,\qquad h^{\textcolor[rgb]{1,0,0}{\#}}({\bf u},{\bf u})\equiv& \prod_{\substack{u_i,u_j\in {\bf u}\\i\textcolor[rgb]{1,0,0}{\#}j}}h(u_i,v_j)\,,
     \end{aligned}
     \eeq 
     where $\textcolor[rgb]{1,0,0}{\#}$ can be either $<$, $>$ or $\neq$.
In addition to these two contributions, the integrand contains the propagation factor $e^{-E(v)l_B}$ and the measure $\mu(v^{\gamma})$ whose explicit expressions can be found in Appendix \ref{FHandWCE}.

 \begin{figure}[t]
\begin{center}
\vspace{-2cm}
\includegraphics[scale=.6]{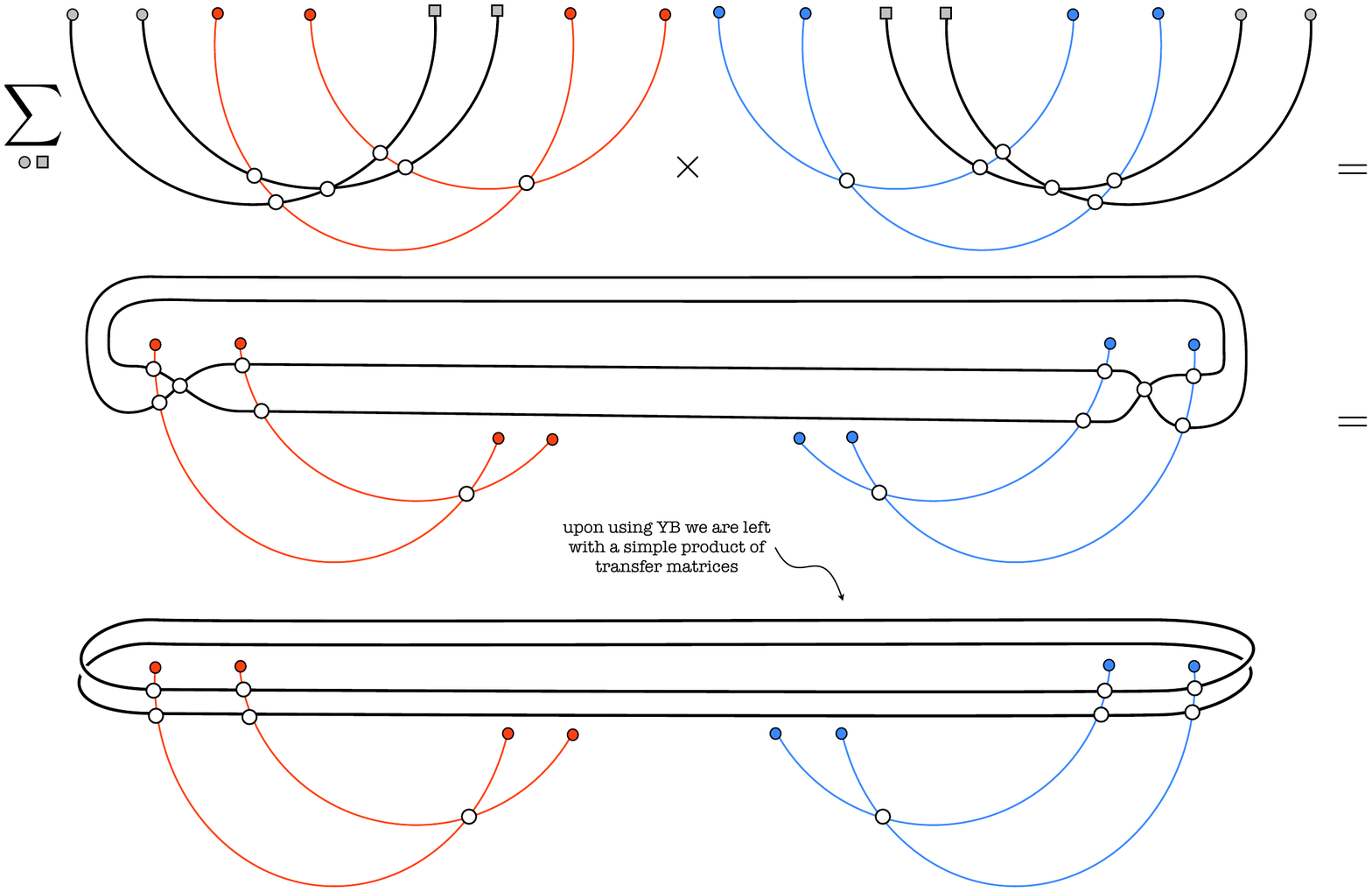}
\end{center}
\vspace{-1.5cm}
\caption{\normalsize Matrix part for two mirror excitations. Owing to the Yang-Baxter relation, the matrix part can be factorized into two $\mathfrak{psu}(2|2)$ transfer matrices. The interaction between mirror particles only appears in the dynamical factors.} \la{YB2} \vspace{-.2cm} 
\end{figure}

For general multi-particle wrappings, the result again consists of the part coming from summation over flavours (the matrix part), the constant of proportionality determined by the configuration with $D$'s and $\bar{D}$'s (the dynamical part), the propagation factors and the measures. Importantly, thanks to the Yang-Baxter relation, the matrix part can always be written as a product of transfer matrices as illustrated in figure \ref{YB2}. By contrast, the dynamical part is more nontrivial since each term in the sum for the asymptotic structure constant will receive different corrections from particles on adjacent edges. 
 
To see this explicitly, let us take a close look at the dynamical part. The simplest way to compute this is to put mirror particles on the top edges of the hexagons as shown in figure \ref{movingthings}. Since the first hexagon only contains ``longitudinal derivatives'' $D$'s, it gives rise to the factorized dynamical factors,
\beq\la{fromfirstgeneral}
 \begin{aligned}
 &\hspace{-0.5cm}h^{<}(\wB^{-3\gamma},\wB^{-3\gamma})h^{<}(\wL^{-5\gamma},\wL^{-5\gamma})h^{<}(\wR^{-\gamma},\wR^{-\gamma}) h(\wR^{-\gamma},\wB^{-3\gamma})h(\wR^{-\gamma},\wL^{-5\gamma})h(\wB^{-3\gamma},\wL^{-5\gamma})\\
& \times h( \alpha, \wB^{-3\gamma})h(\alpha, \wL^{-5\gamma})h(\alpha, \wR^{-\gamma})\,.
\end{aligned}
\eeq
As for the second hexagon, it is convenient to perform mirror transformations and bring the particles to the right hand side of the top edge (see figure \ref{movingthings}). After this manipulation, mirror particles become $D$'s and the result is again given by a factorized expression,
\beq\la{fromsecondgeneral}
 \begin{aligned}
 &\hspace{-0.5cm}h^{>}(\wB^{-3\gamma},\wB^{-3\gamma})h^{>}(\wL^{-\gamma},\wL^{-\gamma})h^{>}(\wR^{-5\gamma},\wR^{-5\gamma}) h(\wL^{-\gamma},\wB^{-3\gamma})h(\wL^{-\gamma},\wR^{-5\gamma})h(\wB^{-3\gamma},\wR^{-5\gamma})\\
& \times h( \bar{\alpha}, \wB^{-3\gamma})h(\bar{\alpha}, \wL^{-\gamma})h(\bar{\alpha}, \wR^{-5\gamma})\,.
\end{aligned}
\eeq

\begin{figure}[t]
\begin{center}
\includegraphics[scale=.6]{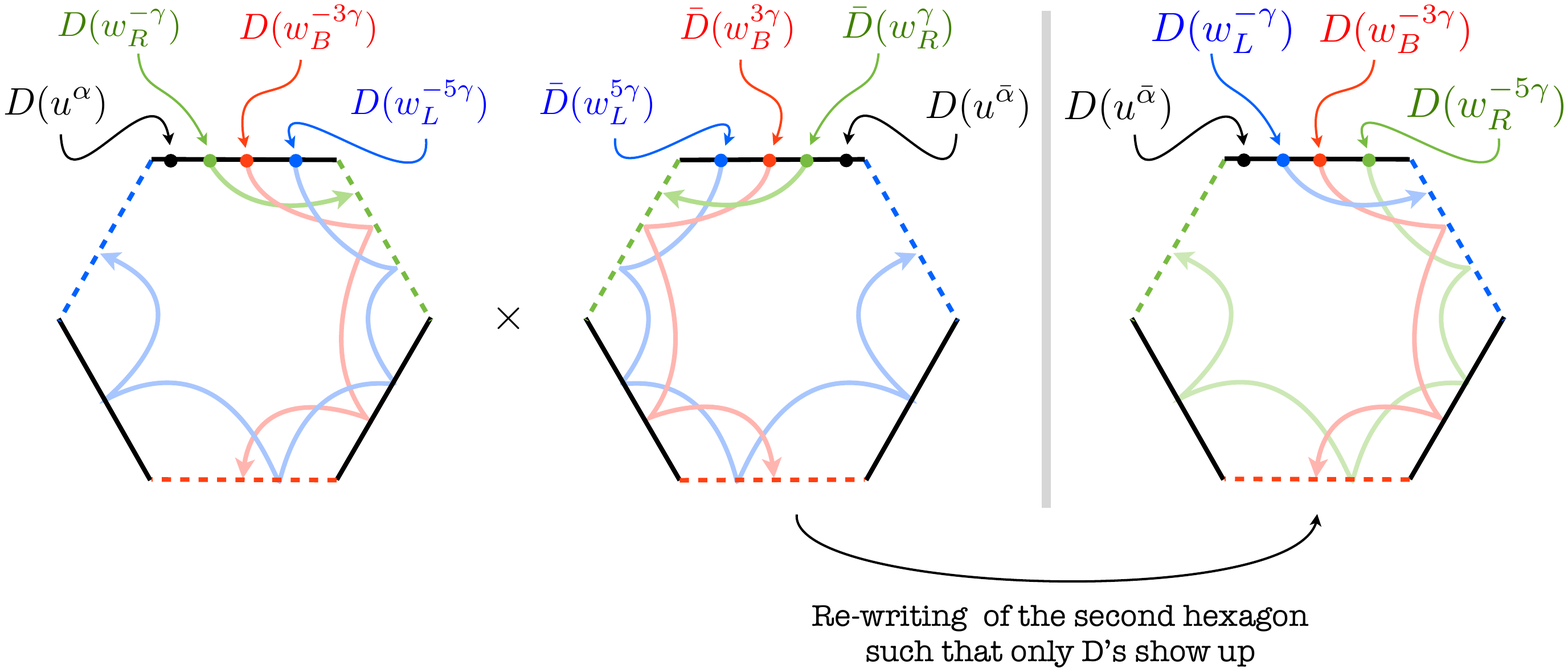}
\end{center}
\vspace{-5.5cm}
\caption{\normalsize The configuration for general wrapping corrections. One can decorate mirror edges by particles by putting them on the top edge as shown above. To evaluate the contribution from the second hexagon, it is convenient to utilize the mirror transformation and convert all the particles into $D$. After this manipulation, it becomes clear that the dynamical part is given simply by factorized expressions (\ref{fromfirstgeneral}) and (\ref{fromsecondgeneral}).} \la{movingthings} \vspace{-.2cm} 
\end{figure}

Having determined the dynamical part, it is now straightforward to write down the full expression by putting together the matrix part, the measures and the propagation factors\footnote{We denoted the rapidities of particles on the left adjacent edge in blue, the right adjacent edge in green and the bottom edge in red to make it clear which part comes from which wrapping.}:
\beq\label{multi-integrand}
\begin{aligned}
{\tt Integrand}=&\,\mu (\bottom{\wB^{\gamma}})\mu (\lefta{\wL^{\gamma}})\mu (\righta{\wR^{\gamma}}) e^{-E (\bottom{\wB})l_B}e^{-E (\lefta{\wL}) l_L}e^{-E (\righta{\wR}) l_R} T(\bottom{\wB^{\gamma}})T(\lefta{\wL^{-\gamma}})T(\righta{\wR^{-\gamma}})\\
&\times h^{\neq} (\bottom{\wB^{\gamma}},\bottom{\wB^{\gamma}}) h^{\neq} (\lefta{\wL^{\gamma}},\lefta{\wL^{\gamma}})h^{\neq} (\righta{\wR^{\gamma}},\righta{\wR^{\gamma}})h(\lefta{\wL^{-\gamma}},\righta{\wR^{-5\gamma}})h(\righta{\wR^{-\gamma}},\lefta{\wL^{-5\gamma}})\\
&\times h({\bf u},\bottom{\wB^{-3\gamma}}) \sum_{\alpha\cup \bar{\alpha}={\bf u}}(-1)^{|\bar{\alpha}|}e^{ip_{\bar{\alpha}}l_R}\frac{h(\alpha,\lefta{\wL^{-5\gamma}})h(\alpha,\righta{\wR^{-\gamma}})h(\bar{\alpha},\lefta{\wL^{-\gamma}})h(\bar{\alpha},\righta{\wR^{-5\gamma}})}{h(\alpha,\bar{\alpha})}
\end{aligned}
\eeq
Here we used the properties of $h$ and $T$, $h(u^{2\gamma},v^{2\gamma})=h(u,v)$, $h(u^{4\gamma},v)=1/h(v,u)$ and $T(u^{4\gamma})=T(u)$, to simplify the expression. To generalize (\ref{multi-integrand}) to the cases including bound states, one just needs to substitute $h(u,v)$, $\mu(u)$, $T(u)$ and $E(u)$ with their bound-state counterparts, which are given in Appendix \ref{FHandWCE}.

An important feature of (\ref{multi-integrand}) is that the matrix part is given by a product of transfer matrices. This makes it easy to take into account cancellation due to supersymmetry and enables us to estimate the coupling-constant dependence quite accurately. As will be derived in Appendix \ref{FHandWCE}, various factors in (\ref{multi-integrand}) scale as
\beq
\begin{aligned}
&h(w_i^{\gamma},w_j^{\gamma})\sim \mu(w^{\gamma}) \sim e^{-E(w)}\sim O(g^2)\,,\qquad h(w_i^{-\gamma},w_j^{-5\gamma})\sim h(w_i^{-5\gamma},w_j^{-\gamma})\sim O(g^{-2})\,,
\end{aligned}
\eeq
{while all other factors are $O(1)$.}
From this, one can determine the coupling-constant dependence of the integrand as
\beq
{\tt Integrand}\sim O\left( g^{2\left[n_Bl_B+n_Ll_L+n_Rl_R+n_B^2+(n_L-n_R)^2+n_L+n_R\right]}\right)\,,
\eeq
{as anticipated in figure \ref{hexagon}.}

{Another interesting feature of (\ref{multi-integrand}) is the absence of any interaction between particles on the bottom edge and particles on the adjacent edges. In particular, there are no kinematical poles between particles in the bottom and particles in one of the adjacent edges (in contradistinction with the case where two particles face each other in opposing adjacent edges). The absence of such poles has a natural interpretation: As mentioned in section \ref{conclusion}, such a pole corresponds to a physical process where a particle goes around one of the operators and the residue will be given by the $\mathfrak{psu}(2|2)^2$ transfer matrix of the operator. However, since the operator is BPS in this case, the transfer matrix vanishes and so does the residue. In other words, the absence of poles is another manifestation of the SUSY cancellation.  }

\section{Transfer Matrices} \la{Tap}
Transfer matrices in symmetric and antisymmetric representations were computed in~\cite{Beisert:2006qh}. 
The asymptotic transfer matrices in the anti-symmetric representation are the ones relevant for our analysis and take the form
\beq
T_{a}(u) = (-1)^a\sum_{n=-1}^{1}(3n^2-2)\prod_{m=0}^{n}\frac{R^{(+)}(u^{[2m-a]})}{R^{(-)}(u^{[2m-a]})} \sum_{j=\frac{2-a}{2}}^{\frac{a-2n}{2}}\prod_{k=j+n}^{\frac{a-2}{2}}\frac{R^{(+)}(u^{[2n-2k]})B^{(+)}(u^{[-2k]})}{R^{(-)}(u^{[2n-2k]})B^{(-)}(u^{[-2k]})}\, ,
\eeq
where
\beq
R^{(\pm)}(u) = \prod_{j}(x(u)-x^{\mp}_{j})\, , \,\,\, B^{(\pm)}(u) = \prod_{j}(\frac{1}{x(u)}-x^{\mp}_{j})\, .
\eeq
Its mirror version $u\rightarrow u^{\pm \gamma}$ is directly obtained by crossing $x^{[\pm a]}\rightarrow 1/x^{[\pm a]}$, while keeping fixed the remaining Zhukowsky variables. 
Of course, this transformation does not commute with perturbation theory. Thus, for perturbative studies it is convenient to have independent expansions of these transfer matrices after various mirror rotations. We have
\beqa
T_a(u^{-\gamma}) \!\!\! &=&\!\!\! -g^2 \gamma \frac{(-1)^{a}}{2\,Q^{[-a+1]}}\sum_{k=-\frac{a-1}{2}}^{\frac{a-1}{2}} \frac{Q^{[2k]}}{u^{[2k+1]}u^{[2k-1]}} + O(g^4)\, , \la{T1}\\
T_a(u)\!\!\!  &=&\!\!\! \frac{(-1)^a}{Q^{[-a-1]}}\Big(  Q^{[-a-1]}-Q^{[-a+1]} -  i g^2 \frac{\gamma}{2} \Big(\frac{2\,Q^{[-a+1]}}{u^{[-a]}}-\!\!\! \sum\limits_{k=-\frac{a-1}{2}}^{\frac{a-1}{2}} \frac{i \,Q^{[2k]}}{u^{[2k+1]}u^{[2k-1]}}\Big)+O(g^4)\Big) \, ,  \la{Tuseless} \\
T_a(u^{+\gamma})\!\!\!  &=&\!\!\! \frac{(-1)^a}{Q^{[-a-1]}}\Big(1+\frac{i}{2}g^2 \frac{\gamma}{u^{[-a]}}\Big) \Big(  Q^{[a+1]}+Q^{[-a-1]}-Q^{[a-1]}-Q^{[-a+1]}  \la{T2} \\
&+&\!\!\!   i g^2 \frac{\gamma}{2}\Big(\frac{Q^{[a+1]}}{u^{[a]}}-\frac{Q^{[-a-1]}}{u^{[-a]}}+\Big(\frac{Q^{[a-1]}}{u^{[a-2]}}-\frac{Q^{[-a+1]}}{u^{[-a+2]}}\Big) \delta_{a\neq 1}+\!\!\! \sum\limits_{k=-\frac{a-3}{2}}^{\frac{a-3}{2}} \frac{i \,Q^{[2k]}}{u^{[2k+1]}u^{[2k-1]}}\Big)+  O(g^4) \Big)\, ,\nn
\eeqa
where $\gamma=  \sum\limits_{j=1}^S \frac{2}{u_j^2+\frac{1}{4}}$ is the one loop anomalous dimension and $Q(u)\equiv \prod\limits_{j=1}^S (u-u_j)$ is the Baxter polynomial. We are using the standard short-hand notation $f^{[a]}=f(u+i a/2)$ and in particular $u^{[a]}=u+ia/2$. 

The leading order limit of (\ref{T2}) was used in \cite{C123Paper}; the expansion in this channel is relevant for mirror excitations in the bottom mirror edge. 
The expression in (\ref{T1}) is relevant for mirror excitations in one of the adjacent edges. It is also this expansion which is relevant for the study of Luscher correction in the spectrum problem. The middle line (\ref{Tuseless}) for a physical rapidity $u$ is presented here for completeness but is not being used. 

\section{Fused Hexagons and Weak Coupling Expansions}\label{FHandWCE}
Here we summarize various weak coupling expansions of measures and fused pentagon transitions $$h_a(u,v) = \prod\limits_{k=-\frac{a-1}{2}}^{\frac{a-1}{2}} h(u^{[2k]},v)$$ 
relevant for the analysis in this paper. It is convenient to split the pentagon transitions into its symmetric and anti-symmetric part (since the latter is just the well studied S-matrix):
\beq
h_a(u,v)   h_a(v,u) = p_a(u,v) \,, \qquad  h_a(u,v)/h_a(v,u) = S_a(u,v) \,.
\eeq
The S-matrix reads
\beq
S_a(u,v)= \frac{1}{\sigma_a^2(u,v)}  \, \frac{(u-v+i\frac{a-1}{2})(u-v+i\frac{a+1}{2})}{(u-v-i\frac{a-1}{2})(u-v-i\frac{a+1}{2})}\, \prod_{k=-\frac{a-1}{2}}^{\frac{a-1}{2}} \(\frac{1-\frac{1}{y^+x^{[-2k-1]}}}{1-\frac{1}{y^-x^{[+2k+1]}}}\)^{\!2} \,,
\eeq
where $\sigma_a(u,v)=e^{i \chi(u^{[a]},v^+)+i\chi(u^{[-a]},v^-)-i \chi(u^{[-a]},v^+)-i\chi(u^{[a]},v^-)}$ is the (fused) BES dressing phase~\cite{BES}. The product factor $p_a$ is considerably simpler since the dressing phase drops out in this case and the fusion is also particularly simple, leading to a simple expression purely in terms of Zhukowsky variables, 
\beq
p_a(u,v)  
= \frac{(u-v)^2+\frac{(a-1)^2}{4}}{(u-v)^2+\frac{(a+1)^2}{4}} \( \frac{1-\frac{1}{y^-x^{[+a]}}}{1-\frac{1}{y^-x^{[-a]}}}\frac{1-\frac{1}{y^+x^{[-a]}}}{1-\frac{1}{y^+x^{[+a]}}}\)^2 \,.
\eeq
These results can now be straightforwardly expanded in perturbation theory. For illustration, we provide here some explicit results to leading and sub-leading order at weak coupling
\vskip-\parskip
 {\footnotesize{
\beqa
\!\!\!\!\!\!\!\!\!\!\!\!S_a(u,v)\!\!\!\!  &=&\!\!\!\!   \frac{(u^{[+a-1]}-v)(u^{[+a+1]}-v)}{(u^{[-a+1]}-v)(u^{[-a-1]}-v)} \Big( 1- \tfrac{2g^2}{i} \big( \tfrac{1}{v^+}H(\tfrac{u^{[a]}}{-i})-\tfrac{1}{v^+}H(\tfrac{u^{[-a]}}{-i})+\tfrac{1}{v^-} H(\tfrac{u^{[-a]}}{i})-\tfrac{1}{v^-} H(\tfrac{u^{[a]}}{i}) \big)+\dots \Big)\, , \nn \\
\!\!\!\!\!\!\!\!\!\!\!\!S_a(u^\gamma,v) \!\!\!\! &=&\! \!\!\!\frac{(u^{[+a-1]}-v)(g y^{-})^2}{(u^{[-a+1]}-v)(u^{[-a-1]}-v)(u^{[a+1]}-v)} \Big(1-\frac{g^2}{v^+v^-}\Big(\frac{2i}{u^{[a]}}+\frac{4uv}{u^{[a]}u^{[-a]}}-H(\tfrac{u^{[-a]}}{i})-H(\tfrac{u^{[a]}}{i})-H(\tfrac{u^{[-a]}}{-i})-H(\tfrac{u^{[a]}}{-i}) \Big)\Big)\, , \nn
\eeqa
}}with $H(n)$ the harmonic number, and 
\beqa
&&\!\!\!\!\!\!\!\!\!\!\!\!p_a(u,v) =\frac{1}{p_a(u^{2\gamma},v)}= \frac{(u-v)^2+\frac{(a-1)^2}{4}}{(u-v)^2+\frac{(a+1)^2}{4}}  \(1- \frac{2 a g^2}{(u^2+\frac{a^2}{4})(v^2+\frac{1}{4})}+\dots\)\, , \\
&&\!\!\!\!\!\!\!\!\!\!\!\!p_a(u^\gamma,v)=\frac{1}{p_a(u^{-\gamma},v)}  =\Big(\frac{v^+}{v^-}\Big)^{\!2}\, \frac{u^{[+a+1]}-v}{u^{[-a-1]}-v}\frac{u^{[-a+1]}-v}{u^{[+a-1]}-v}  \( 1+ \frac{i g^2(u+a^2 v+4 u^2 v+4 u v^2)}{(u^2+\frac{a^2}{4})(v^2+\frac{1}{4})^2}+\dots \)\, . \nn
\eeqa
Combining these expansions (with sub-leading terms included of course) with the important relation $h(u^{4\gamma},v)=1/h(v,u)$, we can easily reproduce any of the weak coupling expansions used in this paper. Finally we have the fused measures
\beqa\label{eq:measuregamma}
\mu_a(u)&=&\frac{a(x^{[+a]}x^{[-a]})^2}{g^2(x^{[+a]}-x^{[-a]})^2((x^{[+a]})^2-1)(1-(x^{[-a]})^{2})}=\frac{1}{a}-\frac{a g^2}{\left(\frac{a^2}{4}+u^2\right)^2}+O(g^4)\, , \\
\mu_a(u^\gamma)&=&\frac{a(x^{[+a]}x^{[-a]})^2}{g^2(x^{[+a]}x^{[-a]}-1)^2((x^{[+a]})^2-1)((x^{[-a]})^2-1)} 
=\frac{a g^2}{\left(\frac{a^2}{4}+u^2\right)^2}-\frac{a g^4 \left(a^2-8 u^2\right)}{\left(\frac{a^2}{4}+u^2\right)^4} +O\left(g^6\right)\, .\nn
\eeqa

\section{Transfer Matrix from Sum over Partitions}\la{Tfromsum}
In this appendix, we will show that the integrand for a mirror particle on the adjacent edge with the bridge size $\ell =1$ coincides with the transfer matrix $T_{a}(v^{\gamma})$ at the leading order in the weak coupling expansion.  

As discussed in section \ref{adjacent}, the integrand for a mirror particle on the left adjacent edge is given by
\beq
\mathcal{A}_{\rm adjacent}(v)=\sum_{\alpha\cup \bar{\alpha}={\bf u}} (-1)^{|\alpha|} \prod_{j\in \bar{\alpha}} (e^{ip_{j}\ell}h_a(u_j,v^{-\gamma})h_a(v^{-\gamma},u_j))\!\! \prod_{i\in \alpha, j\in \bar{\alpha}}  \frac{1}{h_{ij}}  \, .
 \eeq
By setting $\ell=1$ and expanding the terms in the sum at weak coupling, we obtain
\beq
\mathcal{A}_{\rm adjacent}(v)=\mathbb{A}(v^{[1-a]},v^{[1+a]})+O(g^2)\,,
\eeq
where the function $\mathbb{A}$ is given by the following sum over partitions:
\beq\la{bbA}
\mathbb{A}(v,w) = \sum_{\alpha\cup \bar{\alpha}={\bf u}} (-1)^{|\bar{\alpha}|} \prod_{\bar{u}\in \bar{\alpha}}\frac{\bar{u}-i/2}{\bar{u}+i/2}\frac{(v-\bar{u}-i)(w-\bar{u}-i)}{(v-\bar{u})(w-\bar{u})} \prod_{\substack{u\in\alpha\\\bar{u}\in \bar{\alpha}}} \frac{u-\bar{u}-i}{u-\bar{u}}\, .
\eeq
The goal of this appendix is to prove the following interesting identity between $\mathbb{A}$ and the leading order $T_a(v^{-\gamma})$,
\beq\la{toprove}
\begin{aligned}
\mathbb{A}(v^{[1-a]},v^{[1+a]})&= \frac{(-1)^ai^{|{\bf u}|}|{\bf u}|!}{-g^2 \gamma}\frac{v^{[a]}v^{[-a]}}{Q^{[1+a]}(v)} T_a (v^{-\gamma})\\
&= \frac{i^{|{\bf u}|}|{\bf u}|!v^{[a]}v^{[-a]}}{2\,Q^{[1+a]}Q^{[1-a]}}\sum_{k=-\frac{a-1}{2}}^{\frac{a-1}{2}} \frac{Q^{[2k]}}{v^{[2k+1]}v^{[2k-1]}}\,.
\end{aligned}
\eeq

As a first step, let us prove the following identity, which is valid for any rapidity set ${\bf w}$ with the cardinality $|{\bf w}|>1$:
\beq\la{impossible}
\mathcal{A}_{\rm su2}({\bf w})\equiv\sum_{\alpha\cup \bar{\alpha}={\bf w}} (-1)^{|\bar{\alpha}|} \prod_{\bar{w}\in \bar{\alpha}}\frac{\bar{w}-i/2}{\bar{w}+i/2} \prod_{\substack{w\in\alpha\\\bar{w}\in \bar{\alpha}}} \frac{w-\bar{w}-i}{w-\bar{w}}=0\,.
\eeq
To show this, we just need to notice that $\mathcal{A}_{\rm su2}({\bf w})$ is identical to the scalar product between the vacuum descendant and general off-shell Bethe states in the SU(2) spin chain with length 1. For the SU(2) spin chain, it is clearly impossible to put more magnons than the length of the chain (see Appendix C of \cite{tailoringIII}). This immediately leads to the identity (\ref{impossible}).

Now, using the definition of $\mathbb{A}$, one can show by straightforward computation that\footnote{{The differential operator used here appeared in \cite{Ivan} in a different context.}}
\beq
\begin{aligned}
&\frac{1}{(s-t)Q(s)Q(t)}\left(1-\frac{s-i/2}{s+i/2}e^{i\partial_{s}}\right)\left(1-\frac{t-i/2}{t+i/2}e^{i\partial_{t}}\right)(s-t)Q(s)Q(t)\mathbb{A}(s,t)
\end{aligned}
\eeq
coincides with $\mathcal{A}_{\rm su2}({\bf u}\cup s\cup t)$ and therefore must vanish. 
Then, setting $s$ and $t$ to be $v^{[1-a]}$ and $v^{[1+a]}$, we get the following functional relation:
\beq\la{functional}
\begin{aligned}
&\mathbb{A}(v^{[1-a]},v^{[1+a]})+\frac{v^{[a]}v^{[-a]}Q^{[3+a]}(v)Q^{[3-a]}(v)}{v^{[2+a]}v^{[2-a]}Q^{[1+a]}(v)Q^{[1-a]}(v)}\mathbb{A}(v^{[3-a]},v^{[3+a]})\\
&=-\frac{(1-a)v^{[-a]}Q^{[3-a]}(v)}{av^{[2-a]}Q^{[1-a]}(v)}\mathbb{A}(v^{[3-a]},v^{[1+a]})+\frac{(1+a)v^{[a]}Q^{[3+a]}(v)}{av^{[2+a]}Q^{[1+a]}(v)}\mathbb{A}(v^{[1-a]},v^{[3+a]})\,.
\end{aligned}
\eeq
In particular for $a=1$, the relation reads
\beq\label{fundamental relation1}
\mathbb{A}(v,v^{[2]})+\frac{v^{[-1]}Q^{[4]}(v)}{v^{[3]}Q(v)}\mathbb{A}(v^{[2]},v^{[4]})=\frac{2v^{[1]}Q^{[4]}(v)}{v^{[3]}Q^{[2]}(v)}\mathbb{A}(v,v^{[4]})\,.
\eeq
Using (\ref{functional}) recursively, one can express $\mathbb{A}(v^{[1-a]},v^{[1+a]})$ in terms of the fundamental one, $\mathbb{A}(v,v^{[2]})$. Furthermore, one can verify that the right hand side of (\ref{toprove}) satisfies the relation (\ref{functional}). Thus, to prove (\ref{toprove}), we only need to show it for $a=1$; the rest follows from (\ref{functional}).

To prove (\ref{toprove}) for $a=1$, we use the following properties, which characterize $\mathbb{A}(v,v^{[2]})$ uniquely:
\begin{enumerate}
\item $\left.\mathbb{A}(v,v^{[2]})\right|_{{\bf u}=\varnothing}=1$.
\item $\mathbb{A}(v,v^{[2]})$ is a rational function whose denominator is given by $Q(-i/2) Q^{[2]}(v)$.
\item $\displaystyle\left.\mathbb{A}(v,v^{[2]})\right|_{{\bf u}}\quad \overset{u_i\to\infty}{\longrightarrow} \quad \left.-\frac{i|{\bf u}|}{u_i}\mathbb{A}(v,v^{[2]})\right|_{{\bf u}\backslash u_i} $
\item $ \displaystyle {\rm Res}_{u_i=i/2}\mathbb{A}(v,v^{[2]})=i\left(\prod_{j\neq i}\frac{(u_j-i/2)}{(u_j+i/2)}\right)\frac{(v-i/2)}{(v+3i/2)} Z[v|{\bf u}\backslash u_i]$, where $Z[v|{\bf u}]$ is given by
\beq\la{Zvu}
Z[v|{\bf u}] = \sum_{\alpha\cup \bar{\alpha}={\bf u}} (-1)^{|\bar{\alpha}|} \prod_{\bar{u}\in \bar{\alpha}}\frac{v-\bar{u}-i}{v-\bar{u}+i} \prod_{\substack{u\in\alpha\\\bar{u}\in \bar{\alpha}}} \frac{u-\bar{u}-i}{u-\bar{u}}
\eeq
\end{enumerate}
The properties $1$, $3$ and $4$ can be straightforwardly shown from the definition of $\mathbb{A}$. To prove the second one, we need to show that $\mathbb{A}(v,v^{[2]})$ does not have a pole when two rapidities $u_i$ and $u_j$ coincide. This follows from the fact that the residues of such a pole for $\{u_i\in \alpha ,u_j\in \bar{\alpha}\}$ and $\{u_i\in \alpha ,u_j\in \bar{\alpha}\}$ have the opposite sign and therefore they disappear after summation. These properties allow us to determine $\mathbb{A}(v,v^{[2]})$ explicitly once $Z[v|{\bf u}]$ is computed. Now, to compute $Z[v|{\bf u}]$, we use the following relation, which can be deduced directly from (\ref{Zvu}):
\beq\la{limitZ}
Z[v|{\bf u}]\quad \overset{u_i\to\infty}{\longrightarrow} \quad -\frac{i(|{\bf u}|+1)}{u_i}Z[v|{\bf u}\backslash u_i]
 \eeq 
From (\ref{limitZ}) and $Z[v|\varnothing]=1$, we can conclude that $Z[v|{\bf u}]$ is given by
\beq
Z[v|{\bf u}]=\frac{i^{|{\bf u}|}(|{\bf u}|+1)!}{Q^{[2]}(v)} \,.
\eeq

Having determined $Z[v|{\bf u}]$, one can now use the properties listed above and prove by mathematical induction that $\mathbb{A}(v,v^{[2]})$ is given by
\beq\la{AbyBaxter}
\mathbb{A}(v,v^{[2]})= \frac{i^{|{\bf u}|-1}|{\bf u}|!}{Q^{[2]}(v)}\left[ \left(1-\frac{Q(i/2)}{Q(-i/2)}\right)v+ \left(1+\frac{Q(i/2)}{Q(-i/2)}\right)\frac{i}{2}\right]\,.
\eeq
Finally using the zero-momentum condition $Q(i/2)/Q(-i/2)=1$, we arrive at (\ref{toprove}) for $a=1$. This completes the proof of the formula.

Now, to derive the equality (\ref{Aadjacent}) used in the main text, we also need to know $\mathcal{A}_{\rm asymptotic}$ for $\ell=1$. This can be determined in a similar manner as $Z[v|{\bf u}]$, namely by studying the behavior at $u_i \sim \infty$. The result is
\beq\la{explicitasympt}
\mathcal{A}_{\rm asymptotic} =\frac{i^{|{\bf u}|}|{\bf u}|!}{Q(i/2)}\,.
\eeq
Dividing (\ref{toprove}) by (\ref{explicitasympt}), we obtain the formula (\ref{Aadjacent}).
\section{Harmonic Polylog Technology}\la{HPL}
In this appendix we present an efficient method for computing the wrapping corrections in the opposing channel:
\beq
\frac{\delta\mathcal{A}}{\mathcal{A}}=\int \frac{du}{2\pi}\sum_{a\geq 1}\,\mu_{a}^{\gamma}(u)\bigg(\frac{1}{x^{[+a]}x^{[-a]}}\bigg)^l \frac{(-1)^{a}T_{a}(u^\gamma)}{\prod_{i}h_{D_aD}(u^{\gamma},u_i)}\, .\label{eq:finiteSizeCorrection}
\eeq
Here the transfer matrix $T_{a}(u^\gamma)$ and the measure $\mu^{\gamma}(u)$ are given in (\ref{T2}) and (\ref{eq:measuregamma}) respectively, while the hexagon phase for bound states at weak coupling reads
\vskip-\parskip
{\footnotesize{
\begin{align}
\frac{1}{Q^{[-a-1]}\prod_{i}h_{D_{a}D}(u^{\gamma},u_i)}&=\frac{1}{\prod\limits_{j=1}^S gx_j^{+}}+\gamma  g^2 \frac{H(\frac{-u^{[2-a]}}{i})+H(\frac{-u^{[2+a]}}{i})+H(\frac{u^{[a-2]}}{i})+H(\frac{u^{[-a-2]}}{i})-\frac{2i}{ u^{[-a]}}}{4\prod_{j=1}^S gx_j^{+}}\, ,\nonumber
\end{align}
}}with $\gamma$ the one-loop anomalous dimension, $\sum\limits_{j=1}^S \frac{2}{u_j^2+\frac{1}{4}}$, and $H(x)$ the harmonic sum. 

{The basic idea is to first replace the Baxter polynomials $Q(u)$ with a plane wave $e^{iut}$, compute the integral as a function of $t$ and then perform the differential operator $Q(-i\partial_{t})$ setting $t=0$ in the end. There are two advantages of this method: First, once we compute the integral as a function of $t$, we can straightforwardly generate the results for any spin (namely any Baxter polynomials) simply by differentiation. Second, the plane wave makes the integral more convergent and allows us to compute it simply by taking the residues.}
 
To understand how the method works in practice, let us first rederive the leading order wrapping correction in the opposing channel computed in \cite{C123Paper} for a bridge of size $\ell=1$. At this order, the integrand is given by
\begin{align}
\frac{\delta\mathcal{A}}{\mathcal{A}}\bigg|_{\ell=1} = g^4\sum_{a=1}^{\infty}\int \frac{du}{2\pi }\frac{a}{\big(u^2+\frac{a^2}{4}\big)^3}\frac{Q^{[a+1]}+Q^{[-a-1]}-Q^{[a-1]}-Q^{[1-a]}}{\prod\limits_{j=1}^S gx_j^{+}}+O(g^6).
\end{align}
As noted in \cite{C123Paper}, this integral does not converge well for large spin operators. However, once we replace $Q$ with $e^{iut}$, the integral becomes more convergent and one can compute it simply by closing the contour of integration in the upper half plane (assuming $t\ge 0$) and picking up the residues as follows:
\vskip-\parskip
{\footnotesize{
\begin{align}
\!\!\frac{\delta\mathcal{A}}{\mathcal{A}}\bigg|_{\ell=1} =&\,\frac{Q(-i\partial_{t})}{\prod\limits_{j=1}^S gx_j^{+}} \bigg[\sum_{a=1}^{\infty} \frac{\left(e^t-1\right) e^{-\left(a+\frac{1}{2}\right) t} \left(e^{a t}-1\right) \left(a^2 t^2+6 a t+12\right)}{2 a^4}\bigg]_{t=0}+O(g^6)\nonumber\\
=&\, \frac{g^4Q(-i\partial_{t})}{\prod\limits_{j=1}^S gx_j^{+}}\bigg[\sinh \left({\textstyle{\frac{t}{2}}}\right) \left(H_1^2 \left(6 \zeta _2-H_{10}\right)+6 H_1 \left(H_{110}+2 \zeta _3\right)-12 \left(\zeta _2 H_{11}+H_{1110}\right)\right)\bigg]_{t=0}\!\!\!+O(g^6)\, ,\label{eq:TwoloopsRedo}
\end{align}
}}where the argument of each harmonic polylogarithm $H_{\dots}$ is $1-e^{-t}$. To obtain the expression in the second line, we used the HPL package \cite{Maitre:2005uu}. It is {in fact easy to check} that this result, after differentiation, reproduces the two-loop wrapping correction derived in \cite{C123Paper}. 

With this new method, we now compute the three-loop correction. At three loops, we have to consider the correction coming from Bethe roots in (\ref{eq:TwoloopsRedo}) and also the correction to the integrand (\ref{eq:finiteSizeCorrection}) which are given by
\vskip-\parskip
{\footnotesize{
\begin{align}
&\int \frac{du}{2\pi} \,\frac{\mu_{a}^{\gamma}(u)}{x^{[+a]}x^{[-a]}} \frac{(-1)^{(a)}T_{a}(u^\gamma)}{\prod_{i}h_{D_aD}(u^{\gamma},u_i)}\bigg|_{g^6} =g^6\frac{Q(-i\partial_{t})}{2 \prod\limits_{j=1}^S gx_j^{+}}\bigg[\sinh \big({\textstyle{\frac{t}{2}}}\big) \big[6 \gamma  \zeta _3 H_{11}-2 \gamma  H_{11001}+2 \gamma  H_{10011}+2 \gamma  H_{01101}\nonumber \\
&-2 \gamma  H_{01011}-\gamma  H_{11011}+\gamma  H_{10111}-80 \zeta _3 H_{111}-8 \left(2 H_{111101}-H_{111011}+H_{110111}-2 H_{101111}+40 \zeta _5 H_1\right)\nonumber\\
&-2 \gamma  \zeta _2 \zeta _3+12 \gamma  \zeta _3 H_{01}\big]+\gamma  \cosh \big({\textstyle{\frac{t}{2}}}\big) \left(6 \zeta _3 H_{11}+H_{11101}-H_{11011}+20 \zeta _5\right)\bigg] + (\ref{eq:TwoloopsRedo})+O(g^8).
\end{align}
}}

Let us just point out that all the sums that appear after doing the integral by picking up its residues can also be expressed as harmonic polylogarithms, and the most complicated ones take the following form
\vskip-\parskip
{\footnotesize{
\begin{align}
\sum_{a=1}^{\infty}\frac{z^{a}}{a^{b}}S_{d}(a-1)=H_{\overset{\underbrace{0\dots 01}}{b}\overset{\underbrace{0\dots 01}}{d}}(z).
\end{align}
}}

At three loops we can also consider the finite size correction for a bridge of size $\ell=2$
\vskip-\parskip
{\footnotesize{
\begin{align}
&\frac{\delta\mathcal{A}}{\mathcal{A}}\bigg|_{\ell=2} = g^6\int \frac{du}{2\pi }\sum_{a\ge 1} \frac{a}{\big(u^2+\frac{a^2}{4}\big)^4}\frac{Q^{[a+1]}+Q^{[-a-1]}-Q^{[a-1]}-Q^{[1-a]}}{\prod\limits_{j=1}^S gx_j^{+}}+O(g^8)\\
&=\frac{g^6Q(-i\partial_{t})}{\prod\limits_{j=1}^S gx_j^{+}}\bigg[\frac{ \sinh \left(\frac{t}{2}\right)}{3 } \bigg[(60 H_1 \left(H_{11110}-\zeta _3 H_{11}+\zeta _2 H_{111}+2 \zeta _5\right)-120 \left(H_{111110}-\zeta _3 H_{111}+\zeta _4 H_{11}+\zeta _2 H_{1111}\right)\nonumber\\
&+H_1^3 \left(H_{110}+12 \zeta _3\right)-12 H_1^2 \left(\zeta_2 H_{11}+H_{1110}-5 \zeta _4\right)+H_1^4 \zeta _2\big]\bigg]_{t=0}+O(g^8).\nonumber
\end{align}
}}

{  Interestingly, all the integrals we computed are given in terms of harmonic polylogarithms, 
Riemann zeta functions, $\cosh (\frac{t}{2})$ and $\sinh (\frac{t}{2})$. If this is true in general, one can determine the function of $t$ directly by first constructing an ansatz as linear combination of functions with a given transcendental level and then fix the coefficients by computing a finite number of residues. Such a procedure was indeed very powerful for scattering amplitudes \cite{LanceEtAl1} and should be effective also in this case especially at higher loops.}

\end{document}